\documentclass[twocolumn]{aastex631}

\usepackage{natbib}
\usepackage[utf8]{inputenc}
\usepackage{verbatim}
\usepackage{amsmath}
\usepackage{hyperref}
\usepackage{breakurl}
\usepackage{float}
\usepackage{color}

\usepackage{graphicx}%, floatflt}
\usepackage[normalem]{ulem}

\newcommand{\simgt}{\,\hbox{\lower0.6ex\hbox{$\sim$}\llap{\raise0.6ex\hbox{$>$}}}\,}
\newcommand{\simlt}{\,\hbox{\lower0.6ex\hbox{$\sim$}\llap{\raise0.6ex\hbox{$<$}}}\,}

\newcommand\todo{\color{red}$\square$ \normalcolor}

\begin{document}

\title{On the Origin of the Ancient, Large-Scale Cold Front in the Perseus Cluster of Galaxies}

%\author{Bellomi, E.\altaffilmark{1}, ZuHone, J. A.\altaffilmark{1}, Weinberger, R.\altaffilmark{2}, Walker, S. A.\altaffilmark{3}, Zhuravleva, I. \altaffilmark{4}, Ruszkowski, M. \altaffilmark{5} }

\author{Elena Bellomi}
\affiliation{Center for Astrophysics $\vert$ Harvard \& Smithsonian, 60 Garden St., Cambridge, MA 02138, USA}

\author{John A. ZuHone}
\affiliation{Center for Astrophysics $\vert$ Harvard \& Smithsonian, 60 Garden St., Cambridge, MA 02138, USA}

\author{Rainer Weinberger}
\affiliation{Leibniz Institute for Astrophysics, An der Sternwarte 16, 14482 Potsdam, Germany}

\author{Stephen A. Walker}
\affiliation{Department of Physics and Astronomy, University of Alabama in Huntsville, Huntsville, AL 35899, USA}

\author{Irina Zhuravleva}
\affiliation{Department of Astronomy and Astrophysics, The University of Chicago, Chicago, IL 60637, USA}

\author{Mateusz Ruszkowski}
\affiliation{Department of Astronomy, University of Michigan in Ann Arbor, USA}

\author{Maxim Markevitch}
\affiliation{NASA Goddard Space Flight Center, Code 662, Greenbelt, MD 20771, USA}


\begin{abstract}
The intracluster medium of the Perseus Cluster exhibits spiral-shaped X-ray surface brightness discontinuities known as ``cold fronts'', which simulations indicate are caused by the sloshing motion of the gas after the passage of a subcluster. Recent observations of Perseus have shown that these fronts extend to large radii. In this work, we present simulations of the formation of sloshing cold fronts in Perseus using the AREPO magnetohydrodynamics code, to produce a plausible scenario for the formation of the large front at a radius of 700 kpc. Our simulations explore a range of subcluster masses and impact parameters. We find that low-mass subclusters cannot generate a cold front that can propagate to such a large radius, and that small impact parameters create too much turbulence which leads to the disruption of the cold front before it reaches such a large distance. Subclusters which make only one core passage produce a stable initial front that expands to large radii, but without a second core passage of the subcluster other fronts are not created at a later time in the core region. We find a small range of simulations with subclusters with mass ratios of $R \sim 1:5$ and initial impact parameter of $\theta~\sim~20-25^\circ$ which not only produce the large cold front but a second set in the core region at later times. These simulations indicate that the ``ancient' cold front is $\sim$6-8.5 Gyr old. For the simulations providing the closest match with observations, the subcluster has completely merged into the main cluster.
\end{abstract}
%%%%%%%%%%%%%%%%%%%%%%%%%%%%%%%%%%%%%%%%%%%%%%%%%%%%%%%%%%%%%%%
%%% LEGEND:
%%%%%%%%%%%%%%%%%%%%%%%%%%%%%%%%%%%%%%%%%%%%%%%%%%%%%%%%%%%%%%%
\begin{comment}
\section{notes}
{\color{red}!!} at beginning = copy and paste from other articles\\
\todo = to do\\
The sections starting with ** will be gone\\
- all times have an incertitude of $\pm$0.05 Gyr, because my timestep for the simulations are 0.1Gyr
\end{comment}
%%%%%%%%%%%%%%%%%%%%%%%%%%%%%%%%%%%%%%%%%%%%%%%%%%%%%%%%%%%%%%%
%%%     INTRO                      
%%%%%%%%%%%%%%%%%%%%%%%%%%%%%%%%%%%%%%%%%%%%%%%%%%%%%%%%%%%%%%%
\section{Introduction}
% galaxy clusters 
Galaxy clusters are the largest gravitationally bound objects in the current universe, comprised of galaxies, dark matter (hereafter DM), and a hot, diffuse, and magnetized plasma known as the intracluster medium (hereafter ICM). The ICM is observed in the X-ray band, emitting primarily via thermal bremsstrahlung and collisional ionization processes, and at millimeter wavelengths via the Sunyaev-Zeldovich effect.

Since the process of cosmological structure formation is still ongoing, clusters grow via accretion and mergers with other clusters. Mergers in particular produce very dramatic effects as observed in X-rays, including shock fronts, cold fronts, and indications of gas turbulence via surface brightness (hereafter SB) fluctuations \citep[a recent review of the effects of mergers on the ICM in galaxy clusters as seen in X-rays is presented in][]{zuhone_merger_2022}. These observed features are of primary importance for the investigation of the internal dynamics of clusters and the constraints they may place on the detailed physics of the cluster plasma. 
 
% cold front   	
Cold fronts (hereafter CFs) in particular have attracted significant interest, due to the fact that they are ubiquitous within galaxy clusters \citep{ghizzardi_cold_2010}. CFs were seen in some of the earliest observations of clusters by {\it Chandra}, as sharp SB discontinuities that were initially perceived to be shock fronts \citep{Markevitch_1999,markevitch_chandra_2000,vikhlinin_moving_2001}. However, measurements of the plasma temperature showed that the brighter (and therefore denser) side of the discontinuity is colder, and in most cases the thermal pressure is continuous. These are the characteristics of a contact discontinuity, which can be readily formed when the cold, dense gas of a cluster core is brought into contact with warmer, less dense parts of the ICM by subsonic gas motions. Recent reviews of CFs in merging clusters include \citet{markevitch_shocks_2007,zuhone_cold_2016,zuhone_merger_2022}.

% sloshing CF
One proposed mechanism for forming CFs in the ICM of clusters is that of gas ``sloshing'' in the gravitational potential well of an otherwise relatively relaxed galaxy cluster, which has a very dense, low-temperature core (so-called ``cool-core'' clusters or CCs). In this case, the cold and low-entropy gas from the center of the cluster has been pushed away from the potential minimum, setting off oscillatory gas motions which produce spiral-shaped CFs. It is typically assumed that the gas sloshing is initiated by the passage of a subcluster. This process has been examined in significant detail by a number of simulation works \citep[e.g.][]{ascasibar_origin_2006,zuhone_stirring_2010,roediger_gas_2011,Roediger2012}.
These simulations have shown that CFs, which form near the cluster center slowly, propagate radially outward and are long-lasting, provided they are not destroyed by a subsequent passage of the same subcluster or another merging event \citep[though see][for evidence of the resilience of sloshing CFs to multiple mergers]{Vaezzadeh2022}. Indeed, observations in the last 10 years have shown large-scale CF structures in many galaxy clusters, including Abell 2142 \citep{rossetti_abell_2013}, RXJ2014.8-2430 \citep{walker_2014}, and Abell 1763 \citep{douglass_2018}, which all show CFs at around half of the virial radius.

A paradigmatic example of a cool-core is the Perseus Cluster. Perseus is very bright and nearby, at $z = 0.01756$, and \textit{Chandra} and \textit{XMM-Newton} observations clearly demonstrate the existence of sloshing-type CFs in the core region and beyond \citep{churazov_xmm-newton_2003,fabian_wide_2011,simionescu_large-scale_2012,walker_is_2017}.

Recent deep observations of Perseus have shown that its CFs extend to very large radii. A CF on the eastern side of the cluster at a radius of $\sim$700~kpc (about half the virial radius) was first observed by \textit{XMM-Newton} \citep{simionescu_large-scale_2012}. A deep \textit{Chandra} observation of this large-scale CF by \citet{walker_split_2018} showed that it has a prominent ``hook''-like feature, which was explained by MHD simulations presented in the same work as the effect of fluid instabilities.

%These types of CFs are observed in Perseus \citep{walker_is_2017}, in which the encounter of an off-axis passage of a subcluster transfer angular momentum to the gas in the Perseus gravitational potential, producing spiral-shaped CFs. 

% large-scale CF
    %The sloshing CFs in Perseus extend from the core \citep{walker_is_2017,sanders_measuring_2020} to the cluster outskirt.
    %Using {\it XMM-Newton} and {\it Chandra} datasets, \citep{simionescu_large-scale_2012,walker_split_2018} found a CF on the east side at 700 kpc from the center. This distance is about half of the virial radius of Perseus.

% CF at 1.2 - 1.7 Mpc
Moreover, there is evidence for CFs at even larger distances from the center of the Perseus cluster. \citet{walker_is_2022} observed two additional SB edges at even larger radii, 1.2~Mpc and 1.7~Mpc. They suggested that these contact discontinuities were sloshing cold fronts with an estimated age of $\sim$9~Gyr. However, \citet{zhang_2020_} argued that these contact discontinuities could be generated by a collision between an accretion shock and a ``runaway'' merger shock. Indeed, \citet{zhu_2021} discovered such a large-scale shock near the virial radius in Perseus using \textit{Suzaku} data.

In this work, we used magnetohydrodynamical simulations to study the propagation of CF to large radii in a Perseus-like cluster and we seek to reproduce the large radii CF observed in Perseus at $\sim 700$ kpc.
Previous theoretical studies mostly focused on the CFs of the core region. In this paper, we investigate the evolution of the sloshing CFs that form in the core region and then expand to large radii. %(is that true for elka paper? virgo a2496)
We aim to study the effects of a long evolution on the CF to understand under which conditions the CFs persist. % what happens for long evolution? do CF persist? under which conditions
In this work, we seek to reproduce the large-scale CFs at 700 kpc shown by \citep{walker_split_2018} and address questions such as what is required to generate and sustain a CF that is capable of propagating to such a radius? What characteristics does the merger need to have to prevent the disruption of the CF before it reaches distances from the core?
In the context of a parameter study of the subcluster properties, we do not only aim to determine the shape and location of the CF, but also to place constraints on the possible position and trajectory of the subcluster. This is a separate but important issue since in many sloshing-CF clusters - including Perseus - the identity of the subcluster is not readily apparent and no identifiable structure is visible in the X-ray observations.

    This paper is structured as follows. In Section~\ref{sec:method} we outline the assumed physics and the code used for the simulations, as well as the properties of the different simulations performed in this work. In Sections~\ref{sec:results} and \ref{sec:discussion}, we will present the results of the simulations and make comparisons to the observations of Perseus. In Section~\ref{sec:conclusions} we summarize these results and present our conclusions. Throughout, we assume a flat $\Lambda$CDM cosmology with $h$ = 0.71, $\Omega_m$ = 0.27, and $\Omega_\Lambda$ = 0.73.

%%%%%%%%%%%%%%%%%%%%%%%%%%%%%%%%%%%%%%%%%%%%%%%%%%%%%%%%%%%%%%%
%%%     METHOD                      
%%%%%%%%%%%%%%%%%%%%%%%%%%%%%%%%%%%%%%%%%%%%%%%%%%%%%%%%%%%%%%%
\section{Method}\label{sec:method}

\subsection{Initial Conditions}\label{sec:initial_conditions}
    
In our simulations, the galaxy cluster merger consists of a large, main cluster, and a smaller infalling subcluster set up on a trajectory where it will have a close encounter with the former. The gas in both clusters is modeled as a magnetized, fully ionized ideal fluid with $\gamma = 5/3$ and mean molecular weight $\mu  = 0.6$, which is in hydrostatic equilibrium with a virialized DM halo which dominates the mass of the cluster. We generate the initial conditions for our idealized simulations using the same general method of previous works \citep{ascasibar_origin_2006,zuhone_stirring_2010,zuhone_cold_2016,zuhone_galaxy_2018,zuhone_sloshing_2019}, though we provide a short summary of the main points here.
    
The main cluster is initially located at the center of the simulation domain at rest. For the total density profile of the main cluster, we use a ``super-NFW'' (sNFW) profile \citep{lilley_super-nfw_2018}: 
\begin{equation}\label{eq:density_snfw}
\rho_{\rm sNFW}(r) = \frac{3M}{16\pi{a^3}}\frac{1}{x(1+x)^{5/2}},
\end{equation}
where $M$ and $a$ are the total mass and scale radius of the DM halo, respectively, and we have substituted $x = r/a$. The sNFW mass profile is
\begin{equation} 
M_{\rm sNFW}(r) = M\left[1-\frac{2+3x}{2(1+x)^{3/2}}\right].
\end{equation}
The sNFW profile has the same dependence on radius in the center as the well-known NFW profile \cite{navarro_universal_1997}: 
$\rho \propto r^{-1}$ as $r\rightarrow 0$, and is used here instead because it falls more quickly at large radii, and thus its mass profile converges as $r \rightarrow \infty$.
        
The three-dimensional gas density distribution of the main cluster is modeled by a sum of a $\beta$-model profile \citep{cavaliere_reprint_1976} and the modified $\beta$-model profile from \cite{vikhlinin_chandra_2006}:
\begin{eqnarray}\label{eq:density_beta_vikhlinin2006}
n_e(r) &=& \frac{n_{e,c1}}{\left[1+(r / r_{c1})^{2}\right]^{1.5\beta_1}} \\
\nonumber &+&
\frac{n_{e,c2}}{\left(1+r^{2} / r_{c2}^{2}\right)^{1.5 \beta_2}} \frac{1}{\left(1+r^{\gamma} / r_{s}^{\gamma}\right)^{\varepsilon / 2\gamma}}.
\end{eqnarray}
For the modified $\beta$-model profile (second term in Equation \ref{eq:density_beta_vikhlinin2006}), the $\alpha$ parameter in the original equation which controls the slope at small radii is set to 0 and hence does not appear in this equation. From \ref{eq:density_snfw}-\ref{eq:density_beta_vikhlinin2006}, the temperature and other relevant profiles can be determined by imposing the hydrostatic equilibrium condition.

\begin{figure*}
\centering
\includegraphics[width=0.9\textwidth]{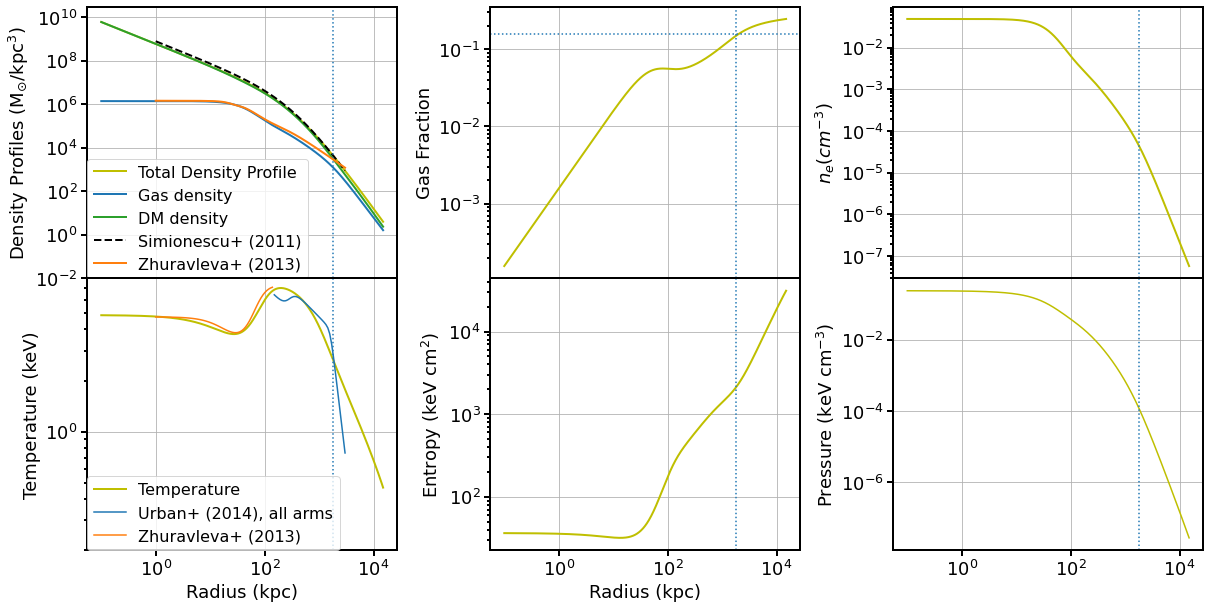}
\caption{Radial profiles of Perseus compared with observations of Perseus. The vertical blue dotted line indicates the virial radius $R_{200c}$ and the horizontal dotted line indicates the cosmological gas fraction 0.156.}
\label{fig:profiles_perseus}

\end{figure*}

To determine the values for the free parameters in Equations \ref{eq:density_snfw}-\ref{eq:density_beta_vikhlinin2006}, we used the observed profiles from \citet{zhuravleva_resonant_2013} and \citet{urban_azimuthally_2014} and chose values for the parameters in Equations \ref{eq:density_snfw}-\ref{eq:density_beta_vikhlinin2006} to closely match these profiles (see Fig.~\ref{fig:profiles_perseus}). We modeled the initial main gas density profile with the parameters specified in Table \ref{tab:perseus_param}.

\begin{table}
\caption{Perseus Model Parameters\label{tab:perseus_param}}
\begin{center}
\begin{tabular}{ l | c  }
\hline\hline
\textbf{Parameter} & \textbf{Value} \\
    \hline\hline
        $M_{200c} $&$ 5.9 \times 10^{14} M_\odot$\\  \hline
        $r_{c1} $&$ 55$ kpc \\	\hline
        $n_{e, c1} $&$ 4.5 \times 10^{-2}$ cm$^{-3}$\\ \hline
        $\beta_1 $&$ 1.2$\\ \hline
        $r_{c2} $&$ 180$ kpc\\ \hline
    $n_{e,c2} $&$ 4\times 10^{-3}$ cm$^{-3}$ \\ \hline
        $r_s $&$ 1800$ kpc\\ \hline
        $\beta_2 $&$ 0.6$\\ \hline
        $\gamma $&$ 3$\\ \hline
        $\varepsilon $&$ 3$\\ 
    \hline\hline
\end{tabular}
\end{center}
\end{table}

 %- subcluster: mass ratio (R), mass-concentration ratio, gasless or gaseous (gas profile)
%- initial trajectory of infalling subcluster + mass ratio + colossus (diemer 2019)
%The two clusters are set up in an isolated computational domain of 40 Mpc, %the merger is simulated in a domain of 40 Mpc and it is characterized by the mass ratio of the main cluster and the infalling subcluter $R = M_{200c,\rm sub}/M_{200c,\rm main}$
	
We model the infalling subcluster in our simulations with a mass imposed by the mass ratio parameter $R = M_{200c,\rm sub}/M_{200c,\rm main}$. The infalling subcluster is initially situated at $d = 3000$~kpc along the $+y$-axis from the main cluster center, with a relative velocity with respect to the main cluster of ${\bf v}_{\rm sub} = v_{\rm sub}( \sin \theta{\bf \hat{x}} -\cos \theta{\bf \hat{y}})$, where $\theta$ is the angle from the $y$-axis, and $v_{\rm sub}$ is the initial speed of the subcluster relative to the main cluster. $R$, $\theta$, and $v_{\rm sub}$ will be varied in different simulations, as detailed below. 

%$v = 1.1 \sqrt{\frac{G \times M_{200c}}{r_{200c}}}$. 
%!!  The initial cluster velocities are chosen so that the total kinetic energy of the system is set to half of its potential energy, under the approximation that the objects are point masses. This results in a relative velocity of the infalling subcluster compared to the center of the main cluster:
%begin{equation}
%    v_{\rm sub} = 1.1 \sqrt{\frac{G M_{\rm main}}{r_{200c}}}
%\end{equation}

%The infalling subcluster gas profile is defined by the second term of eq. \ref{eq:density_beta_vikhlinin2006}, with parameters: 
%Assuming a mass ratio of R, the mass of the infalling subcluster is computed through the concentration-mass relation of Diemer's model \cite{} and assuming a .

We explore simulations where the subcluster can be comprised only of DM, or has a gas component as well. The former type, while physically unrealistic, was motivated in previous works \citep[see especially][]{ascasibar_origin_2006,zuhone_stirring_2010} by a desire to produce smooth and undisrupted CFs; in this work, we show that such an unrealistic restriction is unnecessary given the right choice of parameters. The total mass density profile for the subcluster is also given by Equation \ref{eq:density_snfw}, and (if included) the gas profile is defined as the second term of Equation \ref{eq:density_beta_vikhlinin2006},  with parameters $r_s = 1000$ kpc, $\beta_2 = 2/3$, $\gamma = 3$, and $\varepsilon = 3$. 
The scale density $n_{e,c2}$ is computed assuming a gas fraction defined by Eq. 9 of \cite{vikhlinin_chandracluster_2009}, and the scale radius $r_{c2}$ for the subcluster is a function of $R$ and the concentration parameter: 

        \begin{center}
		\begin{tabular}{ c  }
		    \hline\hline
                $M_{200c, main} = 5.9 \times 10^{14} M_\odot$ \\ \hline
                $r_{c2}($R$~=~1:5) \sim 223$ kpc \\ \hline	
                $r_{c2}($R$~=~1:10) \sim 125$ kpc \\
			\hline\hline	
		\end{tabular}
        \end{center}

The concentration parameter $x$ is computed assuming a concentration-mass relation model from \citet{diemer_accurate_2019} and the Planck 2018 cosmology \citep{collaboration_planck_2020}. %using colossus package in python

We use the results of cosmological simulations to determine the range of initial relative velocities between the main cluster and the subcluster we explore in our parameter space of merger simulations (see Table \ref{table:initial_params}). 
\citet{li_orbital_2020} found that the infall speed of a subhalo follows a nearly lognormal distribution with a skewed tail at large velocities, and is nearly independent of the mass of the subcluster. While the average angle $\theta$ decreases with increasing mass ratio. 
\citet{tormen_1997,vitvitska_2002,li_orbital_2020} pointed out that the most probable velocity is $v_{\rm sub} = 1.1 \sqrt{G M_{200c}/R_{200c}} = 1.1 V_{200c} \sim 1380 $ km~s$^{-1}$. In our study, we also considered higher velocities (i.e. 1500~km~s$^{-1}$ and 2000~km~s$^{-1}$), which correspond to 1.24, 1.65 of the virial circular velocity $V_{\rm circ}$ at the virial radius. 
%Motivated by these results, the velocities we choose in our simulations are 1380, 1500 and 2000 km~s$^{-1}$, which correspond to 1.1, 1.24, 1.65 of the virial circular velocity $V_{\rm circ}$ at the virial radius. The circular velocity is defined as
\begin{equation}
    V_{200c} = \sqrt{GM_{\rm 200c}/R_{\rm 200c}}
\end{equation}
which also corresponds to the infalling velocity at the virial radius.

We also varied the initial angle between the distance vector from the subcluster to the main cluster and their relative velocity vector (see Table \ref{table:initial_params}). Following the study of \citet{li_orbital_2020}, the most probable angle for a merger with a mass ratio of R = 1:5 is around $\theta \sim 30^\circ$, which increases if the mass ratio is smaller. Thus, we decided to explore a range of values for $\theta$ centered around that value of $\theta = 30^\circ$. 
Decreasing this angle is equivalent to decreasing the impact parameter at the beginning of the simulation, which in turn results in a decreased pericenter distance. For some values of the initial velocity and angle, the subcluster will only make one core passage within $\sim$10~Gyr, which is of relevance to the question of where the subcluster that produced sloshing motions can be found in the sky, as we will discuss in Section~\ref{sec:trajectories}.

	\begin{table*}[!ht]
        \caption{Summary of the initial conditions of our parametric study. %The case of a gaseous subcluster with an incident initial angle of 10\degree is not presented here. In addition to this table there set of cases case studied: a gaseous subcluster with an incident angle of 30\degree and intial velocity of 2000 km s$^{-1}$
        } 
        \begin{center}
            \begin{minipage}{0.23\linewidth}
		\centering
		\includegraphics[width=0.67\linewidth]{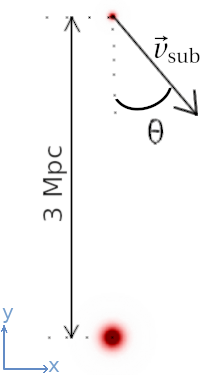}
	\end{minipage}
    \begin{minipage}{0.7\linewidth}
        \begin{tabular}{ l | c }
        \hline\hline
        \textbf{Physical Parameter} & \textbf{Possible Value}s  \\
        \hline\hline	
        gas content of the gas & gaseous / gasless  \\ \hline
        %mass ratio & R =  0.05, 0.1, 0.2, 0.5 \\ \hline
        mass ratio & $R$ =  1:20, 1:10, 1:5, 1:2 \\ \hline
        incident angle & $\theta =$ 10$^\circ$, 20$^\circ$, 30$^\circ$, 40$^\circ$\\ \hline	
        impact parameter & 3000 $\sin(\theta) \sim 529, 1092, 1732, 2517, 3575$ kpc\\ \hline	
        initial velocity & $v = v_{sub} = 1380$, 1500, 2000  km s$^{-1}$\\ \hline
        magnetic field & $\beta = 100, 200, 1000$  \\ 
        \hline\hline
    \end{tabular}
    \label{table:initial_params}
    \end{minipage}
	\end{center}
    \end{table*}

    %%% DISCUSSION OF THE PARAMETER SPACE
    %Is there some range of the physical quantities value that can be excluded by comparing the simulations to the observations?
    %One of the goals of this paper is to study the effect of the different initial parameters. Nevertheless, it is difficult to disentangle the effect of multiple passages of the subcluster to effect due to the initial parameters themselves. For instance, does the CF expand faster if the initial mass ratio is smaller? To answer this type of questions we need to be able to understand if the thing that the CF is faster depends on the mass ratio itself or on the number and angles of the multiple passages of the subcluster (relative to the CF orientation), even though these two last quantities are a consequence of the mass ratio, the initial angle and velocity of the subcluster.
    
\subsection{AREPO code}

The simulations are performed using the moving-mesh cosmological code AREPO code \cite{springel_e_2010} 
which solves the ideal non-radiative magneto-hydrodynamics (MHD) equation employing a finite-volume Godunov method on an unstructured moving mesh, and computes self-gravity via a TreePM solver.  %radiative losses are neglected
    
Each simulation includes $2 \times 10^7$ gas cells and $2 \times 10^7$ DM particles. For each of the initial particle/cell positions, a
random deviate $u = M(<r)/M_{\rm total}$ is uniformly sampled in the range [0, 1] and the mass profile $M(<r)$ for that particular mass type is inverted to
give the radius of the particle/cell from the center of the halo. The 3D position of each particle/cell given its radius is chosen from uniform sampling along the latitudinal and longitudinal directions. 

The gas cells are simulated using the moving-mesh Voronoi tessellation method of AREPO. They are all set initially at the same mass $\sim 10^{7}~M_\odot$, and are allowed to undergo mesh refinement and derefinement as the simulation evolves. The pseudo-Lagrangian nature of the default refinement scheme adopted in AREPO keeps the gas cell mass within a factor of two from a predefined target mass, which is set equal to the initial gas mass of the cells, so that the denser parts of the simulations are better resolved. The initial thermodynamic properties of each gas cell are determined by the profiles from Section~\ref{sec:initial_conditions} and the condition of hydrostatic equilibrium.
    
The bulk of each cluster's mass is made up of the DM particles, which only interact with each other and the gas via gravity. These all have the same mass of $m_{\rm DM} = 4\times10^{7}~M_\odot$. Their initial speeds are determined using the procedure outlined in \cite{Kazantzidis2004}, which computes the particle speed distribution function directly using the method of \cite{Eddington1916}, which assumes the virial equilibrium condition. The velocity vectors for the DM particles are isotropic and determined by choosing random unit vectors in $\Re^3$.
    
All simulations are set within a cubical computational domain of width $L = 40$~Mpc on a side, though for all practical purposes, the region of interest is confined to the inner $\sim$10 Mpc. To ensure that the initial condition is free of spurious gas density and pressure fluctuations from the random nature of the initial conditions, we perform a mesh relaxation step for $\sim 100$ timesteps \citep{springel_e_2010}.

The gas in our simulations is magnetized. For each simulation, we set up the magnetic field on the gas cells following the same approach as in \citet{zuhone_how_2020}. Briefly, we set up a turbulent magnetic field on a uniform grid with a Kolmogorov spectrum which is isotropic in the three spatial directions. The average magnetic field strength is then scaled to be proportional to
the square root of the thermal pressure everywhere in the domain such that $\beta = p_{\rm th}/p_B$ is constant on average (due to the fluctuations in the field, this ratio can only be approximately constant over small spatial regions). The field is transformed to Fourier space to project out the field components that produce $\nabla \cdot {\bf B} \neq 0$, and after transforming back to real space the magnetic field components are then interpolated from this grid onto the cells in each simulation. 

The magnetic fields are evolved on the moving mesh using the Powell approach for divergence cleaning \cite{powell_solution_1999} employed in \cite{pakmor_simulations_2013} and in the IllustrisTNG simulations \cite{marinacci_first_2018}. In this scheme, the divergence of the magnetic field is cleaned through an additional source term in the momentum equation, induction equation, and energy equation. These source terms counteract further growth of local $\nabla\cdot {\bf B}$ errors.

\subsection{Projected Quantities}\label{sec:projected}

In order to compare more directly to the Perseus cluster, we make use of projected maps and images. To produce images of X-ray SB, we compute the X-ray emissivity within the 0.5-7~keV band (in the observer frame) for each gas cell assuming an Astrophysical Plasma Emission Code (APEC) model  \citep{smith_collisional_2001}. Since no metals are included in the simulations, we assume the ICM has a metallicity of $Z = 0.3Z_\odot$ \citep[using relative abundances from][]{solar_metallicity}. For projecting this and other quantities we assume that each gas cell deposits a given volume-weighted quantity onto an image plane perpendicular to the line of sight using a standard cubic SPH smoothing kernel, where the ``smoothing length'' $h_i$ for the kernel for a given particle $i$ is given by
\begin{equation}
h_i = \alpha\left(\frac{3V_i}{4\pi}\right)^{1/3}
\end{equation}
where $V_i$ is the volume of the Voronoi cell and $\alpha = 2$. 
In addition, we create projected temperature maps by weighting each gas cell's temperature by its emission in the same band and projecting along the line of sight. 

Cold fronts are marked out as sharp edges in SB in X-ray observations. This motivates an approach that applies a gradient filter to images to highlight these features. This task can be complicated since X-ray observations can be affected strongly by Poisson noise, especially in low-surface-brightness regions. One way to deal with this complication is to smooth the image before taking its gradient, a technique known as the Gaussian Gradient Magnitude (GGM) filter. This method convolves a noisy 2D image with a Gaussian kernel and then takes the gradient of the image and then computes the magnitude of this gradient field. Mathematically, for an image $I$ and a Gaussian kernel $G$, the edge image $E$ is therefore given by
\begin{equation}
E = |\nabla(G \ast I)| = |(\nabla{G}) \ast I|
\end{equation}
where the last equality follows from the chain rule for convolutions. The size of the Gaussian kernel (the standard deviation $\sigma$) is a free parameter that may be adjusted depending on the noise inherent in the image.

This was introduced by \citet{sanders_detecting_2016} and applied to several clusters in that work and in \citet{swal16}. These works have convincingly demonstrated that this technique may be used as a tool to reveal the existence of edge features not easily seen in SB images (such as cold and shock fronts) which may then be subjected to further analysis. 

Though we do not produce mock X-ray observations of our clusters with the expected Poisson noise from the finite effective area of existing telescopes in this work, the projected SB images we produce still have Poisson noise due to the finite mass resolution of the gas cells. Thus this technique is still useful for displaying the cold and shock fronts produced in our simulations.      
%%%%%%%%%%%%%%%%%%%%%%%%%%%%%%%%%%%%%%%%%%%%%%%%%%%%%%%%%%%%%%%
%%%     RESULTS                      
%%%%%%%%%%%%%%%%%%%%%%%%%%%%%%%%%%%%%%%%%%%%%%%%%%%%%%%%%%%%%%%
\section{Results}\label{sec:results}
    \begin{figure*}% Fig.2
        \centering
        \includegraphics[width=1\textwidth]{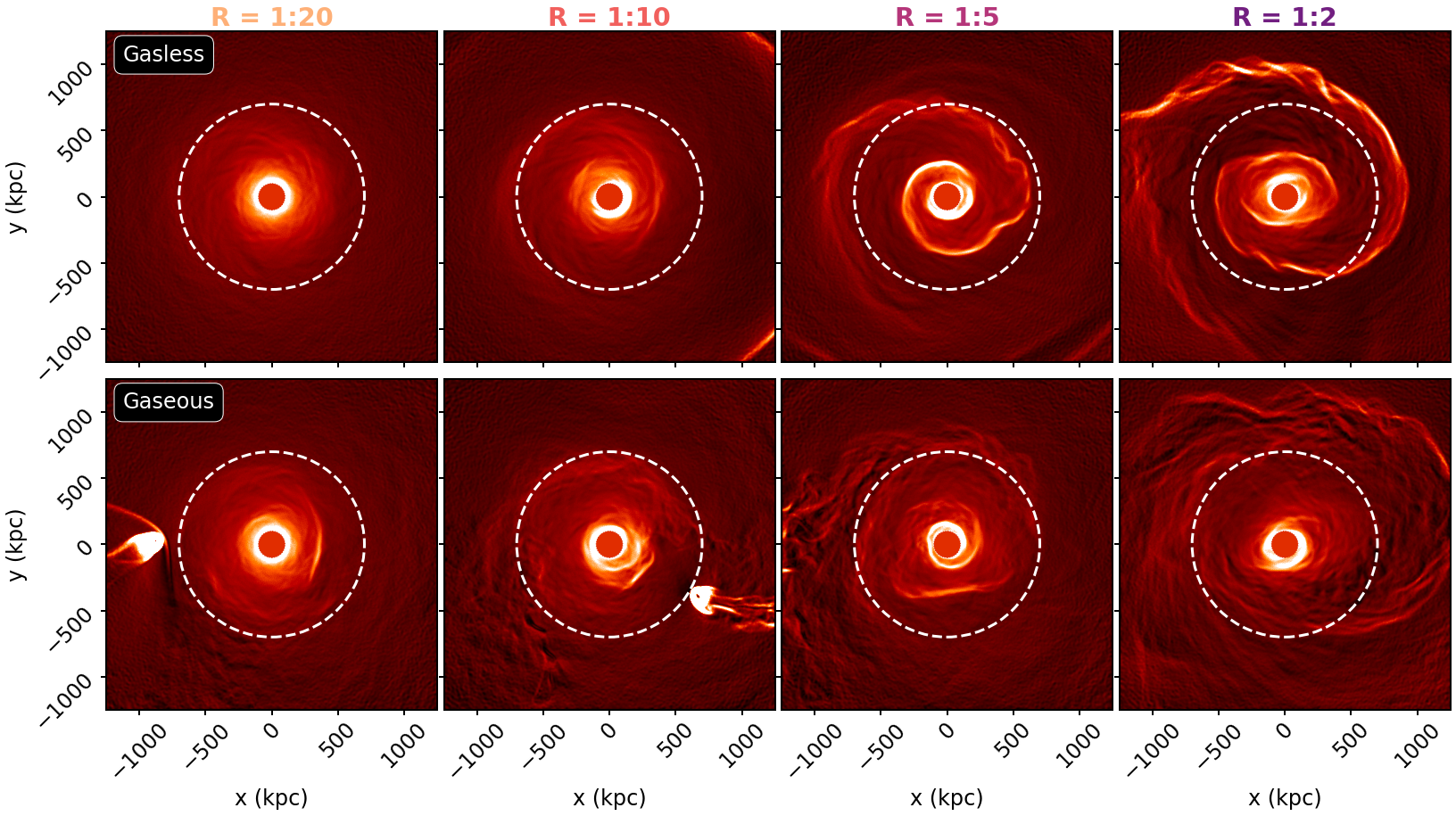}
        \caption{\textit{Different subcluster masses and gas content.} GGM images for the merger simulations with subcluster incoming angle $\theta=20^\circ$, $\beta = 100$ and $v_{\rm sub} = 1380 $ km~s$^{-1}$. Top panels: varying the mass ratio with gasless subclusters. Bottom panels: varying the mass ratio with gaseous subclusters. The white dashed line in each panel marks a circle of radius 700 kpc (the distance from the center at which the ancient CF in Perseus is found). All images are shown at 7.5 Gyr after the first pericenter passage.
        }
        \label{fig:simu_compare_MASS_multiplepassages_GGM}
        
    \end{figure*}
    \begin{figure*}% Fig.3
        \centering
        \includegraphics[width=.8\textwidth]{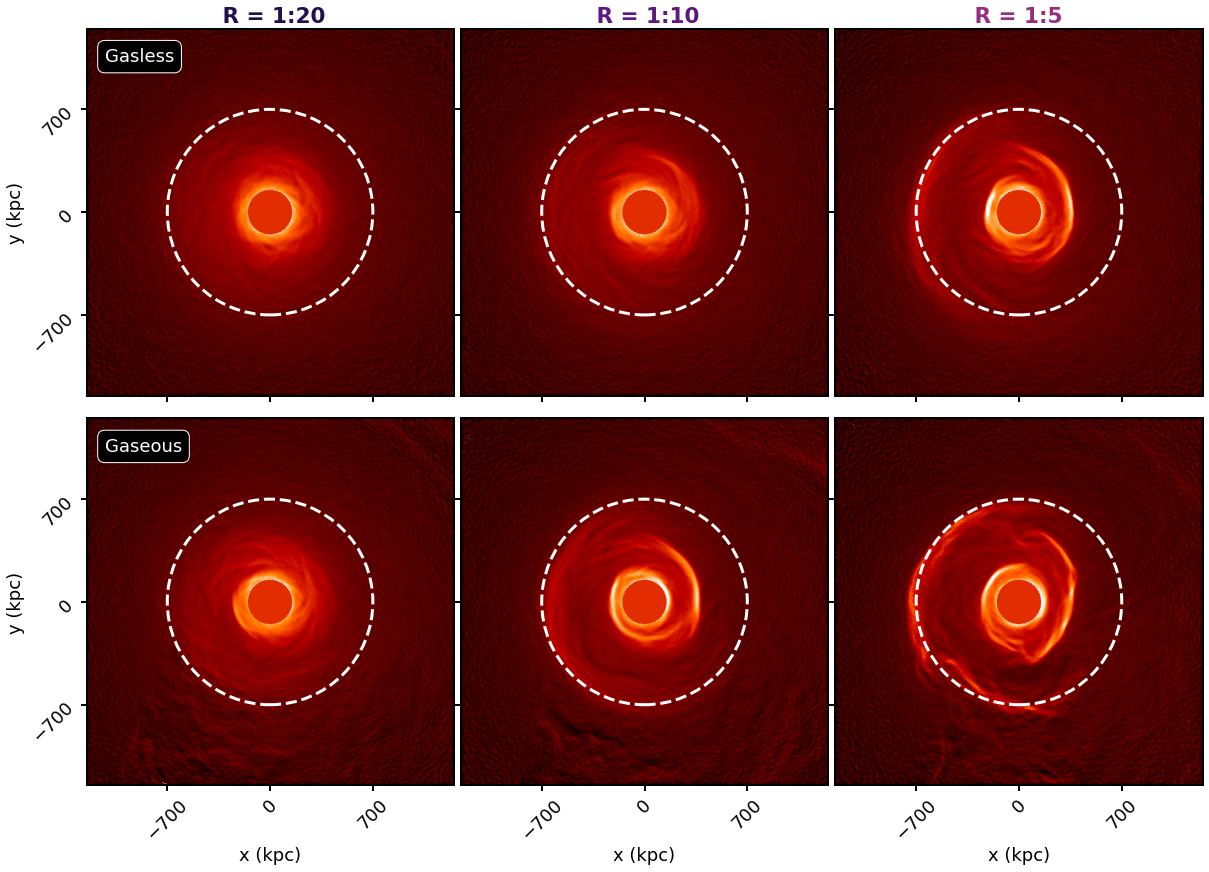}
        \caption{\textit{Different subcluster masses and gas content for a faster encounter.} GGM for the merger simulations with an incident angle $\theta=30^\circ$, initial velocity $v_{\rm sub} = 2000$ km~s$^{-1}$, $\beta = 200$, and mass ratio $R$~=~1:20, 1:10, 1:5. Top panels: varying the mass ratio with gasless subclusters. Bottom panels: varying the mass ratio with gaseous subclusters. The white dashed line in each panel marks a circle of radius 700 kpc (the distance from the center at which the ancient CF in Perseus is found). All images are shown at 7.5 Gyr after the first pericenter passage.
        }
        \label{fig:simu_compare_MASS_singlepassage_GGM}
        
    \end{figure*}
    \begin{figure*}
        \centering
        \includegraphics[width=1\textwidth]{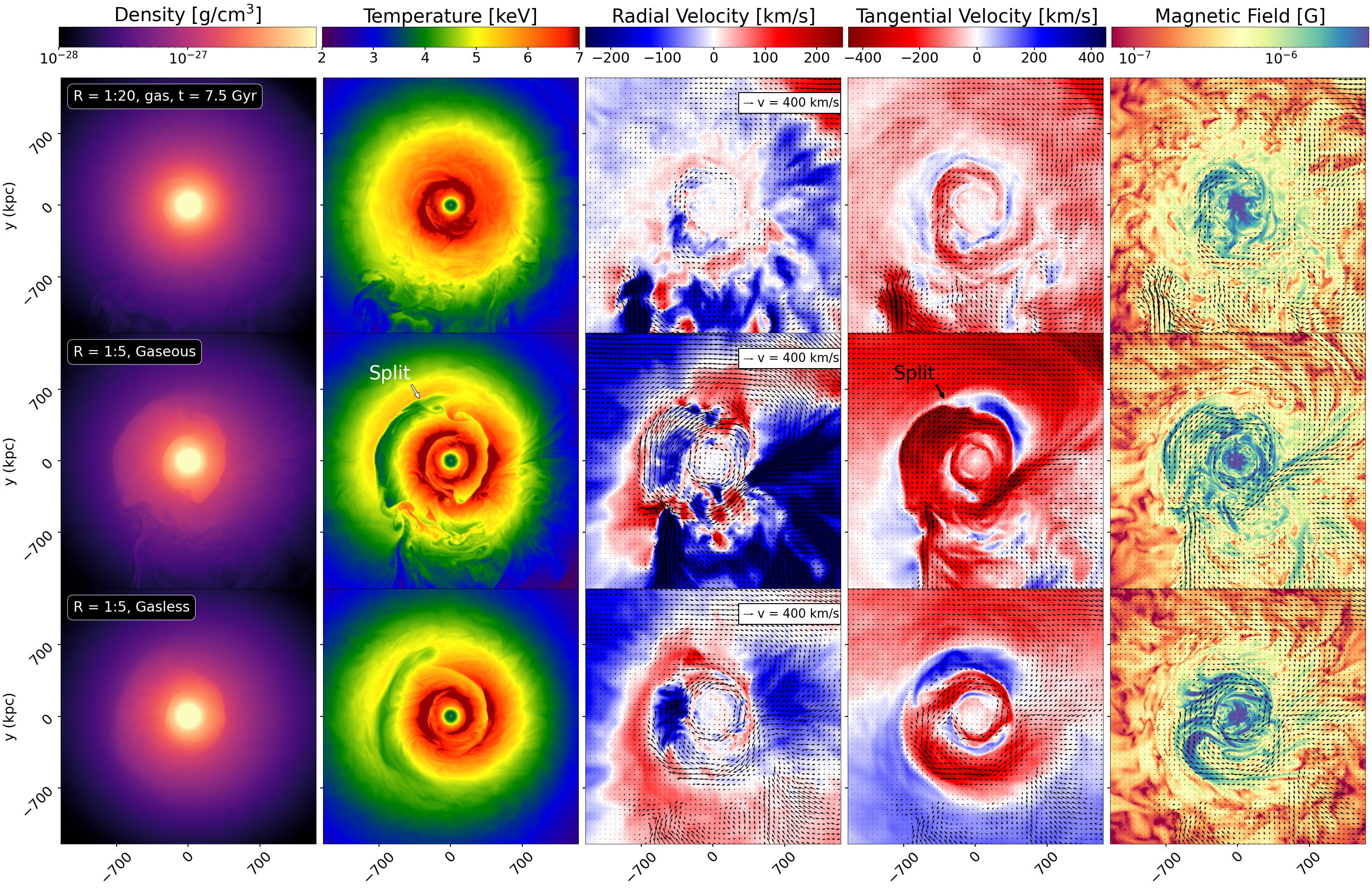}
        \caption{\textit{Different subcluster masses and gas content.} Slices of the density, temperature, radial and tangential velocities, and magnetic field strength for the simulations $R$~=~1:20 gaseous, $R$~=~1:5 gaseous and $R$~=~1:5 gasless, and initial parameters $\theta=30^\circ$, $v_{\rm sub}= 2000$ km~s$^{-1}$, $\beta = 200$. All slices are made through the center of the main cluster, where the center is defined as the minimum of the gravitational potential. All images are shown at 7.5 Gyr after the first pericenter passage. The arrow indicates the split in the CF discussed in Sec.~\ref{sec:split}.
        }
        \label{fig:simu_compare_MASS_singlepassage_slices}
        
    \end{figure*}

    \begin{comment}
    \begin{figure*}
        \centering
        \includegraphics[width=.8\textwidth]{Compare_MASS_gas_beta200_R02_vel2000_multiplot_GGM_mask_2_s25.png}
        \caption{\textit{Different subcluster masses and gas content.} GGM images to highlight the edges of the SB of the simulations of a merger with a subcluster incoming with an incident angle $\theta=30^\circ$, initial velocity $v_{\rm sub}= 2000$ km~s$^{-1}$ (single passage of the subcluster), magnetic parameter $\beta = 200$, and mass ratio $R$~=~1:20, 1:10, 1:5. On the top panels, the case of an encounter with a gasless subcluster and at the bottom in case of a gaseous subcluster.  All images are taken after 2.5 Gyr from the pericenter passage.
        }
        \label{fig:simu_compare_MASS_singlepassage_bowshock}
        
    \end{figure*}
    \end{comment}
    
    % -------------------------
    % velocity + angle Figs
    % -------------------------
    \begin{figure*}[!ht]% Fig.5
        \centering
        \includegraphics[width=1.\textwidth]{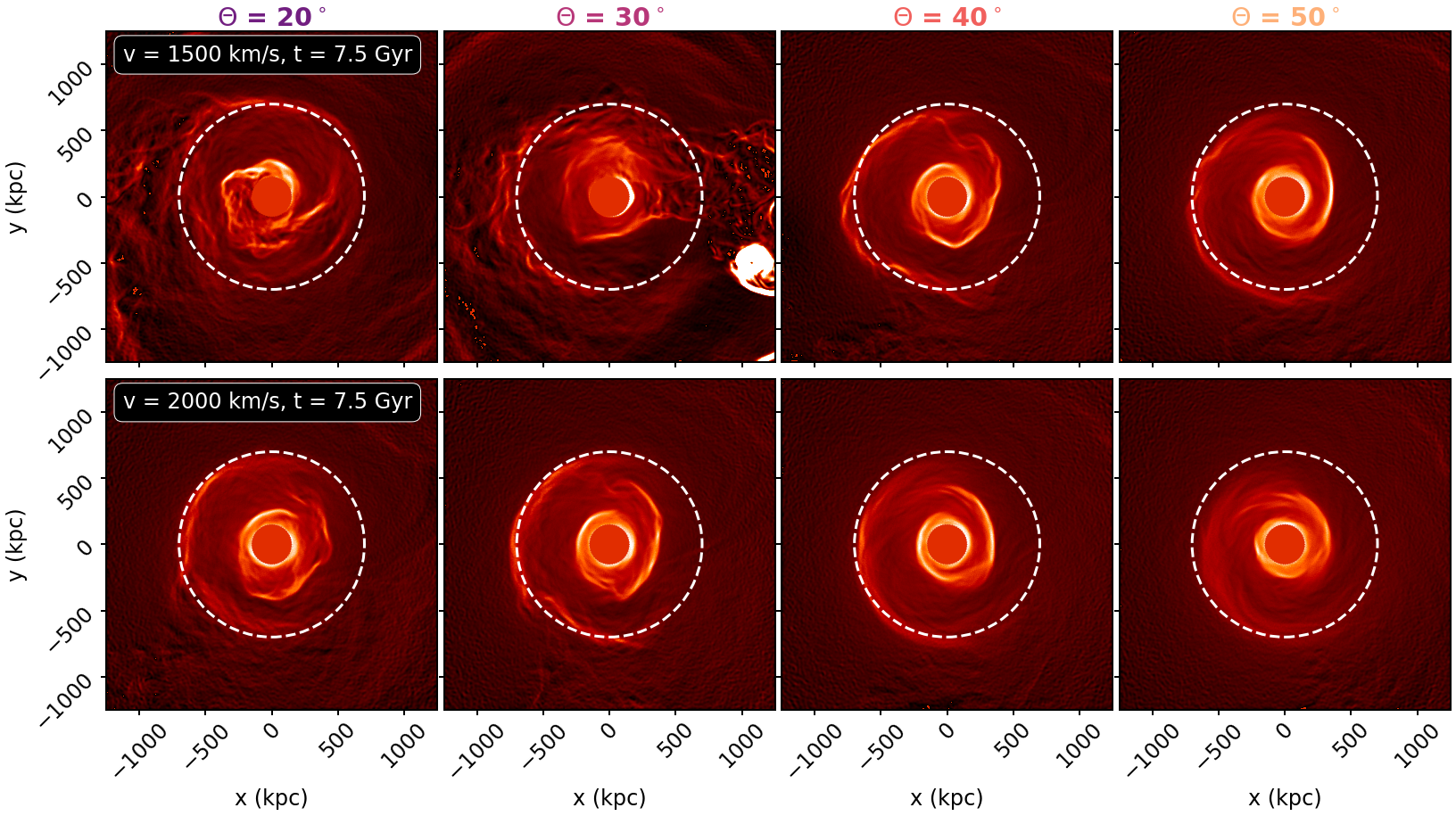}
        \caption{\textit{Different $v_{\rm sub}$ and $\theta$.}
        GGM images for the merger simulations with 2 different initial velocities ($v_{\rm sub}= 1500, 2000$ km~s$^{-1}$) and four different angles ($\theta=20^\circ, 30^\circ, 40^\circ, 50^\circ$). The subcluster is gaseous, the mass ratio is set to $R$~=~1:5 and $\beta = 200$. The white dashed line in each panel marks a circle of radius 700 kpc (the distance from the center at which the ancient CF in Perseus is found). All images are shown at 7.5 Gyr after the first pericenter passage.
        }
        \label{fig:simu_compare_velocity_angle_GGM}
        
    \end{figure*}

    % #############################################
    % LARGE-SCALE CF: MASS, GAS, ANGLE, VELOCITY
    % #############################################
    \subsection{Evolution of the First Cold Fronts}
    
    In this Section, we will determine what parts of our parameter space produce long-lasting CFs that travel out to the large distances seen in observations within a time frame shorter than that between the initial encounter with the subcluster and the current epoch. Our discussion here is broadly consistent with similar discussions of the results of varying the parameter spaces in previous works \citep{ascasibar_origin_2006,zuhone_stirring_2010,roediger_gas_2011,zuhone_sloshing_2011}, though here applied to our specific focus on the Perseus Cluster and on its large-scale cold fronts. 
    %In this section, we will focus on the position and shape of the large-scale CF to constrain the parameter space of the simulation.
    %1. what does it take to have the long-lasting cf that propagates outwards (narrow down parameter space thx to the cf position)     effect of     - changing mass ratio    - whether it has gas of not    - angle    - velocity
    
    \subsubsection{Effect of the subcluster mass and gas content}
    % -------------------------
    % Effect of the MASS
    % -------------------------  
        In Fig. \ref{fig:simu_compare_MASS_multiplepassages_GGM}, we show the GGM images (see Sec. \ref{sec:projected}) for a merger with an incoming subcluster with an initial angle $\theta = 20^\circ$, initial velocity $v_{\rm sub} = 1.1 V_{200c} \sim 1380 $ km~s$^{-1}$, and varying mass of the subcluster by changing the mass ratio parameter $R$ (see Table \ref{table:initial_params}). We also explore the effect of gasless and gaseous subclusters for every value of the mass ratio. % and beta = 100
        All GGM images in this figure are shown 7.5 Gyr after pericenter passage of the subcluster. The white dashed circle indicates a distance of 700~kpc from the center of the main cluster (defined as the cluster potential minimum).
        
        Fig. \ref{fig:simu_compare_MASS_multiplepassages_GGM} demonstrates how the SB gradients increase at the CFs with the mass of the subcluster. For encounters with an initial mass ratio $R < 1:5$ (left-most and center-left panels), the outermost cold fronts are almost nonexistent. As the mass ratio increases, the outermost cold fronts also appear at larger radii at the same epoch. Larger subclusters deliver larger accelerations to the main cluster over a larger volume, permitting the formation of large-scale fronts. The effect of including gas in the subcluster (bottom panels) is to deliver an even stronger acceleration to the gas of the main cluster due to the ram pressure from the subcluster's gas \citep[as first noted in][]{ascasibar_origin_2006}.

        In this particular case (i.e. $\theta = 20^\circ$, and $v_{\rm sub} = 1.1 V_{200c} \sim 1380 $ km~s$^{-1}$), the subcluster returns for multiple core passages, perturbing the core and the CFs multiple times. If the subcluster is more massive, the time between such core passages is shorter, due to the increased dynamical friction in these cases (see Section~\ref{sec:trajectories}), resulting in the core being perturbed more frequently. These additional passages can produce a second set of CFs (see Section~\ref{sec:innerCF}) and alter the trajectory of and/or disrupt the first set. The increased disturbance from the multiple core passages more significantly disrupts the CF and reduces the SB jump in the CF in the cases where the subcluster is gaseous over the gasless case. 
    
    %%%% SINGLE PASSAGE
        By contrast, in Fig. \ref{fig:simu_compare_MASS_singlepassage_GGM} we show the effect of different subcluster masses for a simulation of an encounter with a subcluster coming with $\theta = 30^\circ$ and initial velocity $v_{\rm sub} = 1.65V_{200c}\sim2000$~km~s$^{-1}$, where again all images are plotted at 7.5 Gyr after pericenter passage, as in Figure \ref{fig:simu_compare_MASS_multiplepassages_GGM}.
        %%% single passage definition
        In this case, the subcluster does not return to perturb the gas again during the simulation (which was ran for over 14~Gyr). In contrast to the previous case, where the pericentric radius of the first core passage was smaller and the subcluster had multiple core passages, the radius of the outermost CF seems to be only weakly dependent on the mass of the subcluster, though the SB jump across the front is larger for larger mass ratios, as in Fig. \ref{fig:simu_compare_MASS_multiplepassages_GGM}. The SB jump is increased if the subcluster is gaseous, in contrast to the situation in Fig. \ref{fig:simu_compare_MASS_multiplepassages_GGM}. In this case, the gaseous subcluster applies additional acceleration to the cluster core on the first core passage (thus increasing the SB jumps), without having subsequent multiple passages to disrupt, split, or destroy the CFs.
                
        %%% Slices
        To illustrate these points in more detail, in Fig. \ref{fig:simu_compare_MASS_singlepassage_slices} we show slices of the density, temperature, magnetic field (with superposed velocity field vectors), radial velocity field, and tangential velocity field for three different simulations: $R$~=~1:20 gaseous, $R$~=~1:5 gaseous, and $R$~=~1:5 gasless, all three with initial conditions $\theta = 30^\circ$ and initial velocity $v_{\rm sub} = 2000 $ km~s$^{-1}$ (a single-passage case). 
        As already noted above, we can see that in encounters with subclusters of smaller mass, the density and temperature jumps are not as significant. 
 
        The tangential and radial velocities of the gas are also affected by the mass ratio of the merger, as shown in the center and center-right panels of Fig.~\ref{fig:simu_compare_MASS_singlepassage_slices}, as the passage of a heavier subcluster produces faster motions in both components. For this reason, the magnetic field (right-most panels) is also not as strong in the lower-mass cases, since it grows by shear amplification \citep[][see also Section~\ref{sec:effect_beta}]{zuhone_sloshing_2011,brzycki_parameter_2019}.

        Fig.~\ref{fig:simu_compare_MASS_singlepassage_slices} (bottom two rows) also shows the contrast between the two $R$ = 1:5 simulations with and without gas. For the reasons noted above, there is a greater force delivered to the main cluster's core gas when gas is included for the same subcluster mass. This produces cold fronts with larger gradients in density and temperature and drives higher tangential velocities in the sloshing gas. At this epoch, the cold fronts appear at nearly the same radii in either case. However, due to the interaction with the subcluster's gas, the sloshing motions are more turbulent than in the case with a gasless subcluster, and the cold fronts are more disrupted. Inflowing gas streams stripped from the subcluster also penetrate the core region and disrupt the fronts (the most prominent example appearing in the lower-left, south-east region of the bottom two rows of Fig. \ref{fig:simu_compare_MASS_singlepassage_slices}). Such collisions between inflowing gas streams and cold fronts can also trigger Kelvin–Helmholtz instabilities (hereafter KHI) and can create a ``bay'' in the southern part of the CF, in the direction where the stream penetrates it. Hence, this particular bay only exists in the case of a gaseous subcluster in our simulations, though similar features can appear in other simulations where the subclusters are not gaseous \citep{walker_is_2017}.
        
        Since the subcluster gas has its own magnetic field, the result is that the distribution of the magnetic field in the core region is more turbulent than in the gasless subcluster case, and the magnetic field strength is increased along the cold gas streams that are falling into the cold fronts.

    %%% PERTURBATION OF THE CF / INSTABILITIES
    %The CFs are disturbed by multiple types of instabilities. During the subsequent passages, the fluid of the CF - characterized by regions at different densities - is accelerated impulsively by the shock wave that passes by. This sudden acceleration gives rise to the Richtmyer-Meshkov instability that grows with time. Since the shock is more substantial in the case of a gaseous subcluster, the CFs in the gaseous cases are more disrupted / This contributes to increasing the disruption of the gaseous cases.
    %  Multiple passage, passage of a shock on a fluids with different densities (CF). L'instabilità di Richtmyer-Meshkov si verifica quando due fluidi di diversa densità vengono accelerati impulsivamente. Normalmente questo avviene tramite il passaggio di un'onda d'urto. Lo sviluppo dell'instabilità inizia con perturbazioni di piccola ampiezza che inizialmente crescono linearmente nel tempo
    %\todo Birnboim 2010, Roedinger 2011

    \subsubsection{Effect of varying the speed and incident angle of the subcluster}

    In this Section, we explore the effect of varying the initial speed and angle (or equivalently impact parameter, see Table~\ref{table:initial_params}) of the merger while holding the mass ratio fixed. Varying either of these affects the distance between the clusters at the pericenter passage and the amount of time the subcluster spends near the main cluster center. 
    
    The initial speed determines the pericenter distance, and thus how strongly the subcluster perturbs the gas sitting in the main cluster. If the initial speed is smaller, the subcluster has a smaller pericenter distance and this delivers a stronger gravitational perturbation to the main cluster core (see also Sect.~\ref{sec:trajectories}). The subcluster is also more bound to the main cluster, and returns to pass the core earlier. 
    The initial speed does not seem to affect the radial velocity of the CFs resulting from the initial perturbation. However, the initial speed will determine the frequency of subsequent core passages, which will affect the outward propagation of the CFs (see Section~\ref{sec:expansion_speed} and Fig. \ref{fig:CF_speed}). 
    %\todo plot?

    % -------------------------
    % Effect of THETA               
    % ------------------------- 
    Another quantity that determines the pericenter distance is the initial angle. 
    The effect of this initial angle on the cold fronts is shown in Fig. \ref{fig:simu_compare_velocity_angle_GGM}. Fig.~\ref{fig:simu_compare_velocity_angle_GGM} shows simulations with a an initial magnetic field $\beta=200$, a mass ratio of R = 1:5, and two constant incident velocities $v_{\rm sub} = 1500,~2000$ km~s$^{-1}$ (top and bottom panels, respectively). As we noted above, for the initial velocity $v_{\rm sub} = 2000$ km~s$^{-1}$, the subcluster does not make a second pericenter passage within the age of the universe.
    
    When the impact parameter of the subcluster is larger, the passage of the subcluster creates less turbulence and the CFs appear smoother. For the largest angles studied in this paper, the bow shock created by the passage of the subcluster does not greatly affect the main cluster core, which decreases the contrast of the large-radii CFs. When the angle is smaller, not only do CFs show more evidence of KHI and disruption,
    but the whole image shows more variation in SB. In these cases, turbulence is continuously injected by streams of infalling gas and has the effect of disrupting the outer CFs even more.
    Overall, these results show that gaseous subclusters can produce smooth and relatively undisturbed sloshing CFs provided that the subcluster is far enough away while it still has a significant gas mass, obviating the need to produce these features with artificial gasless subclusters in simulations.

%%%%%%%%%%%%%%%%%%%%%%%%%%%%%%%%%%%%%%%%%%%%%%%%%%%%%%%%%%%%%%%%%%%%%%%%%%%%%
    % -------------------------
    % Distance Figs
    % -------------------------   
    \begin{figure*}% Fig.6
        \centering
        \includegraphics[width=0.48\textwidth]{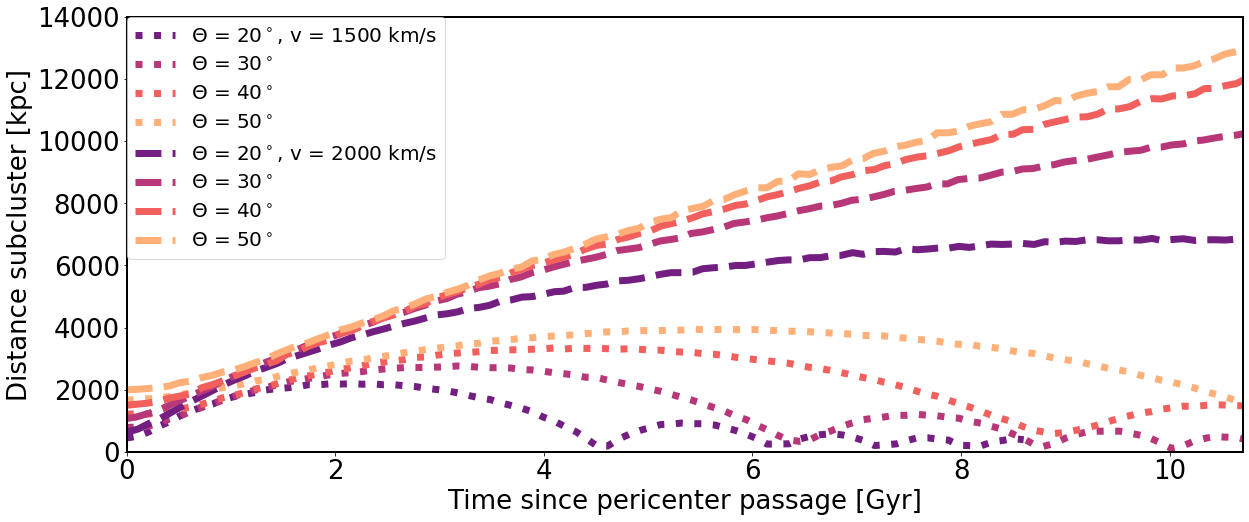}
        \includegraphics[width=0.48\textwidth]{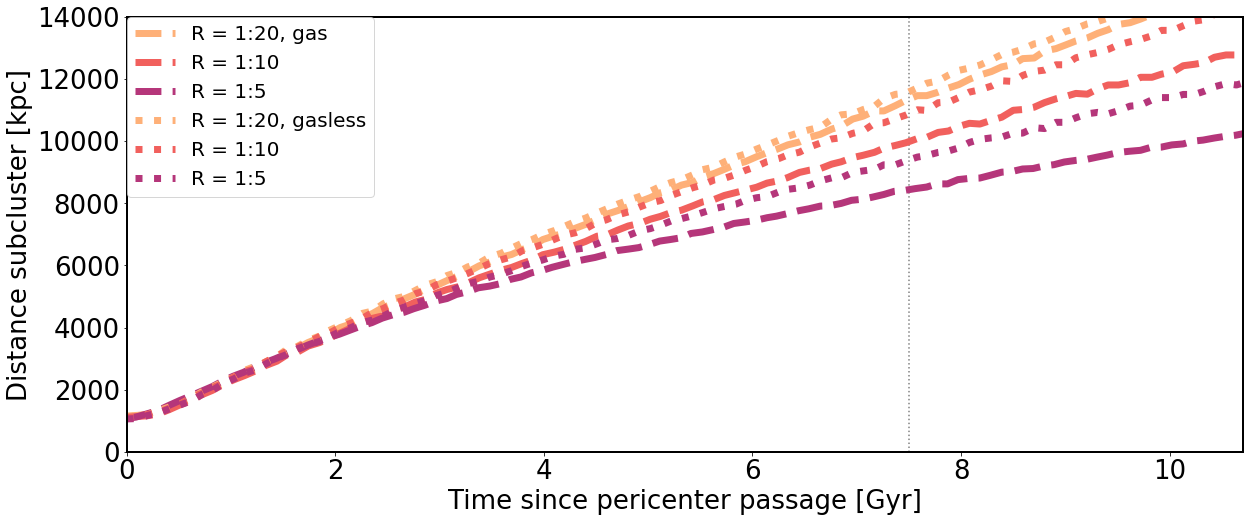}
        \includegraphics[width=0.5\textwidth]{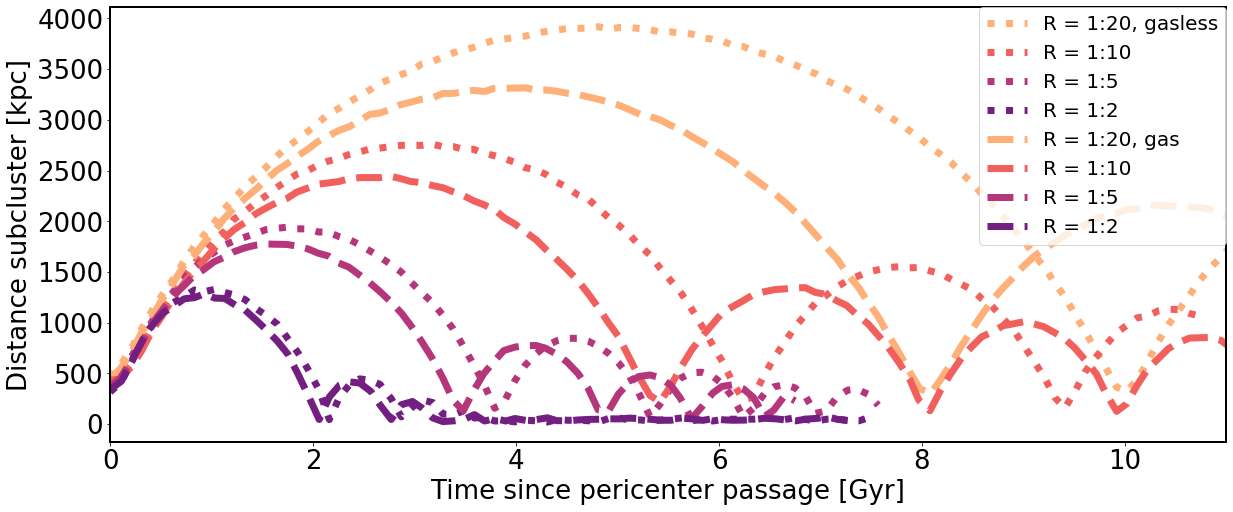}
        \caption{Distance between the center of the main cluster to the center of the subcluster plotted as a function of time for different simulations. The initial time is set to the pericenter passage. 
        \textit{Top Left:} Different initial incident angles and different speeds (see Fig. \ref{fig:simu_compare_velocity_angle_GGM}) 
        for simulations of merger with a gaseous subcluster with mass ratio $R$~=~1:5,  and $\beta = 200$.
        \textit{Top Right:} Different mass and gas content in case of a single passage (see Fig. \ref{fig:simu_compare_MASS_singlepassage_GGM})
        for simulations with $\theta=30^\circ$ and $v_{\rm sub} = 2000 $ km~s$^{-1}$. The dashed lines indicate the gaseous cases while the dotted one indicates the gasless cases.
        \textit{Bottom:} Different mass and gas content in case of multiple passages (see Fig. \ref{fig:simu_compare_MASS_multiplepassages_GGM}) for simulations with $\theta=20^\circ$ and $v_{\rm sub} = 1380 $ km~s$^{-1}$. The dashed lines indicate the gaseous cases while the dotted ones indicate the gasless cases.        }
        \label{fig:simu_compare_distances}
    \end{figure*}
    % -------------------------
    % Trajectory Figs
    % -------------------------   
    \begin{figure*}% Fig.7
        \centering
        \includegraphics[width=1\textwidth]{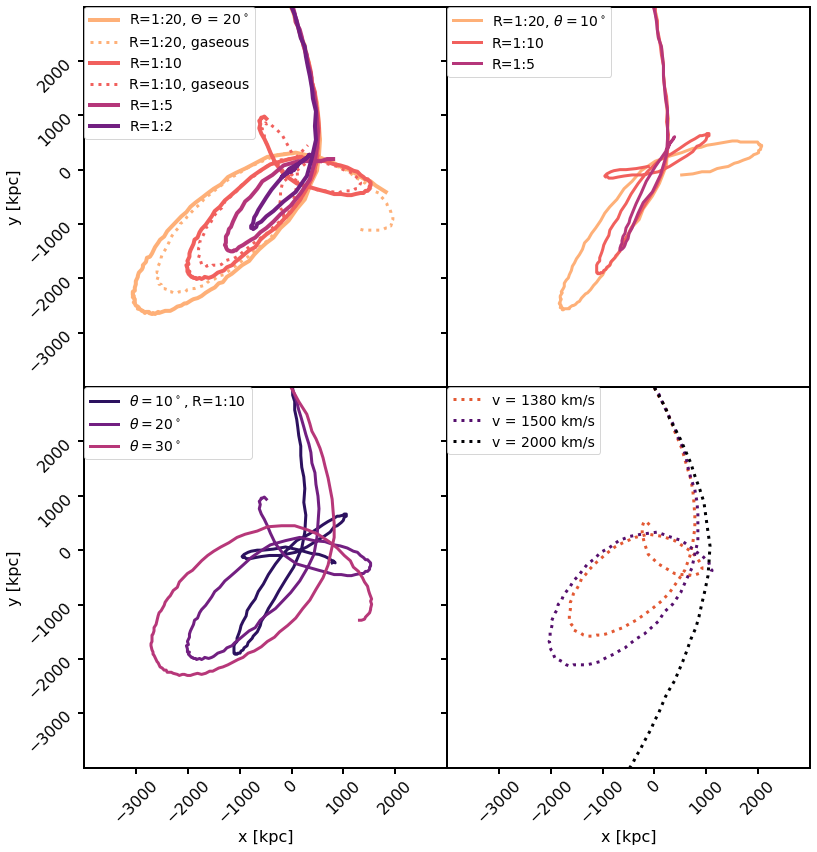}
        \caption{Some examples of the trajectory of the subcluster around the center of the main cluster (the length of each trajectory varies and is chosen for clarity).
        Top-left: Comparison of different trajectories for different mass ratios, and gas and gasless subclusters. The other initial parameters are set to $v_{\rm sub} = 1380 $ km~s$^{-1}$, and $\theta=20^\circ$.
        Top-right: Comparison of different mass ratios for the gasless case and $\theta=10^\circ$ ($v_{\rm sub} = 1380 $ km~s$^{-1}$). Bottom-left:
        Comparison of different initial angles for $R$~=~1:10, gasless subcluster.
        Bottom-right: Comparison for different velocities, $\theta=30^\circ$, $R$~=~1:5, and gaseous subcluster.
        }
        \label{fig:simu_compare_trajectories}
    \end{figure*}

    % #############################################
    % TRAJECTORY
    % #############################################
    \subsection{Subcluster Trajectory}\label{sec:trajectories}
    
    In this section, we discuss in detail how the various merger configurations determine the trajectories and distances of the subcluster to the main cluster, and the sizes and positions of CFs. If the CF characteristics in a simulation can be matched to the observations, this may serve as an indicator of where the subcluster may currently be found, if it has not escaped the main cluster or has already merged with it completely.
    
    %%% DISTANCES: general description
    In Fig. \ref{fig:simu_compare_distances}, we plot the distance between the respective centers (gravitational potential minima) of the subcluster and the main cluster as a function of time from the pericenter passage. In some cases, after the pericenter passage, the subcluster escapes to very large radii and never returns for a second passage within the age of the universe, while in other cases it eventually slows down and then it returns toward the main cluster center. In this paper,  these cases are referred to as ``single-passage'' and ``multiple-passage'' encounters, respectively. 
    
    In the multiple-passage cases, the subcluster increasingly loses DM and/or gas to the main cluster via gravitational capture of DM particles and the ram-pressure stripping of gas as it orbits, and the apocenter of the orbit becomes smaller and smaller with time until the two clusters are completely merged, as this stripped material and the background density from the main cluster exerts dynamical friction on the subcluster \citep{Chandrasekhar1943}. The radii of the apocenters and the changing period of the orbit are a function of the geometry and the mass ratio of the binary merger. In general, the larger the mass ratio or the more ``head-on'' the initial trajectory (smaller initial angle), the more dynamical friction is experienced, the smaller the apocenters and more eccentric the orbits, and the sooner the complete absorption of the subcluster into the main cluster occurs. In minor mergers, the time to complete the merger is generally longer than the age of the universe, see Fig.~\ref{fig:simu_compare_distances}.
        
    In the third panel of Fig. \ref{fig:simu_compare_distances} and in the first panel in Fig.~\ref{fig:simu_compare_trajectories}, we can see the effect of the mass ratio and the gas content on the distance and the trajectory of the subcluster. At a fixed angle, the initial trajectory does not depend much on the mass ratio $R$, since dynamical friction only starts to have a significant impact once the subcluster enters dense regions of the main cluster. Hence, the first pericenter distance is very weakly dependent on the mass of the subcluster. Smaller subclusters result in less dynamical friction since the gravitational focusing effect of matter trailing behind is smaller, so these subclusters reach larger apocenters in shorter periods. The converse is true as the subcluster mass is increased. The number of core passages of the subcluster in a fixed amount of time increases with the mass ratio.

    While the initial trajectory is very weakly affected by the mass of the subcluster, the gravitational interaction between the two clusters greatly affects the trajectory of the subcluster (see Fig. \ref{fig:simu_compare_trajectories}). When the mass ratio increases, the subcluster returns earlier, the orientation of the orbit changes and the orbit is more eccentric. As a consequence, the position and orientation of the subsequent forcings received by the gas in the main cluster are very dependent on the mass ratio (e.g., the trajectory of the subcluster can be parallel to the existing CFs or be perpendicular to them).

    %%% content of gas
    For gaseous subclusters, ram pressure can also have an effect on the subcluster's orbit. For a subcluster of a fixed mass, the ram pressure experienced by the gas of the subcluster results in an additional deceleration on top of that from dynamical friction alone. Though only the gas feels the ram pressure, the material stripped from the subcluster which is dragged behind it exerts an additional gravitational force on the subcluster, slowing it down. As a consequence of this additional drag force, the subcluster orbits have shorter periods and the apocenters are smaller than the gasless cases (see Figs. \ref{fig:simu_compare_distances} and \ref{fig:simu_compare_trajectories}).
    
    %%% initial speed + theta
    The initial speed and the initial incident angle determine the pericenter passage distance. When either decreases, the first passage occurs at a smaller radius and in general the orbit has smaller apocenters and shorter periods.  
    
    In the third panel of Fig. \ref{fig:simu_compare_trajectories}, we plot the trajectories for a subcluster of the same mass ($R$~=~1:10) and different incident angles. These simulations have different impact parameters and also different pericenter radii (examples are shown in the first panel of Fig. \ref{fig:simu_compare_distances} for subcluster of mass ratio $R$~=~1:5). %The smaller the initial angle, the subcluster passes through denser regions, experiencing larger dynamical friction.
    % ECCENTRICITY The semi-major axis is determined entirely by the energy, but the eccentricity depends also on the angular momentum. Both energy and angular momentum are conserved in the two-body problem. This is a many body pb in which the eccentricity change all the time
    The smaller the initial incident angle of the subcluster, the more eccentric the orbit of the subcluster is between the first and second core passages. Also, the subcluster enters denser regions of the core region and experiences more dynamical friction. %This is true because on one hand the impact parameter is smaller, and on the other hand because the subcluster enters a denser region of the main cluster and experience a larger dynamical friction, that slows it down more and make the orbit even more eccentric, as explained before.
    %Larger initial incident angles develop into less eccentric and larger trajectories. 
    %    - different incident angles imply different impact parameters. When the angle is smaller the subcluster pass by denser part of the main cluster, and therefore experience a larger dynamical friction, which affects its later trajectory that appear more eccentric and with a smaller angle compare to the vertical
    % - even mergers that start with a larger incident angles because eventually head-on in future passages, due to dynamical friction, or just spiraling in?

    %%% velocity
    In the fourth panel of Fig. \ref{fig:simu_compare_trajectories}, we plot the trajectories for a subcluster with different initial velocities. The number of orbits and the time for the subcluster to become completely merged into the main cluster is highly dependent on the initial velocity of the subcluster. 
    
    % -------------------------
    % Beta angle FIGS
    % -------------------------
    \begin{figure*} % Fig.8
        \centering
        \includegraphics[width=0.9\textwidth]{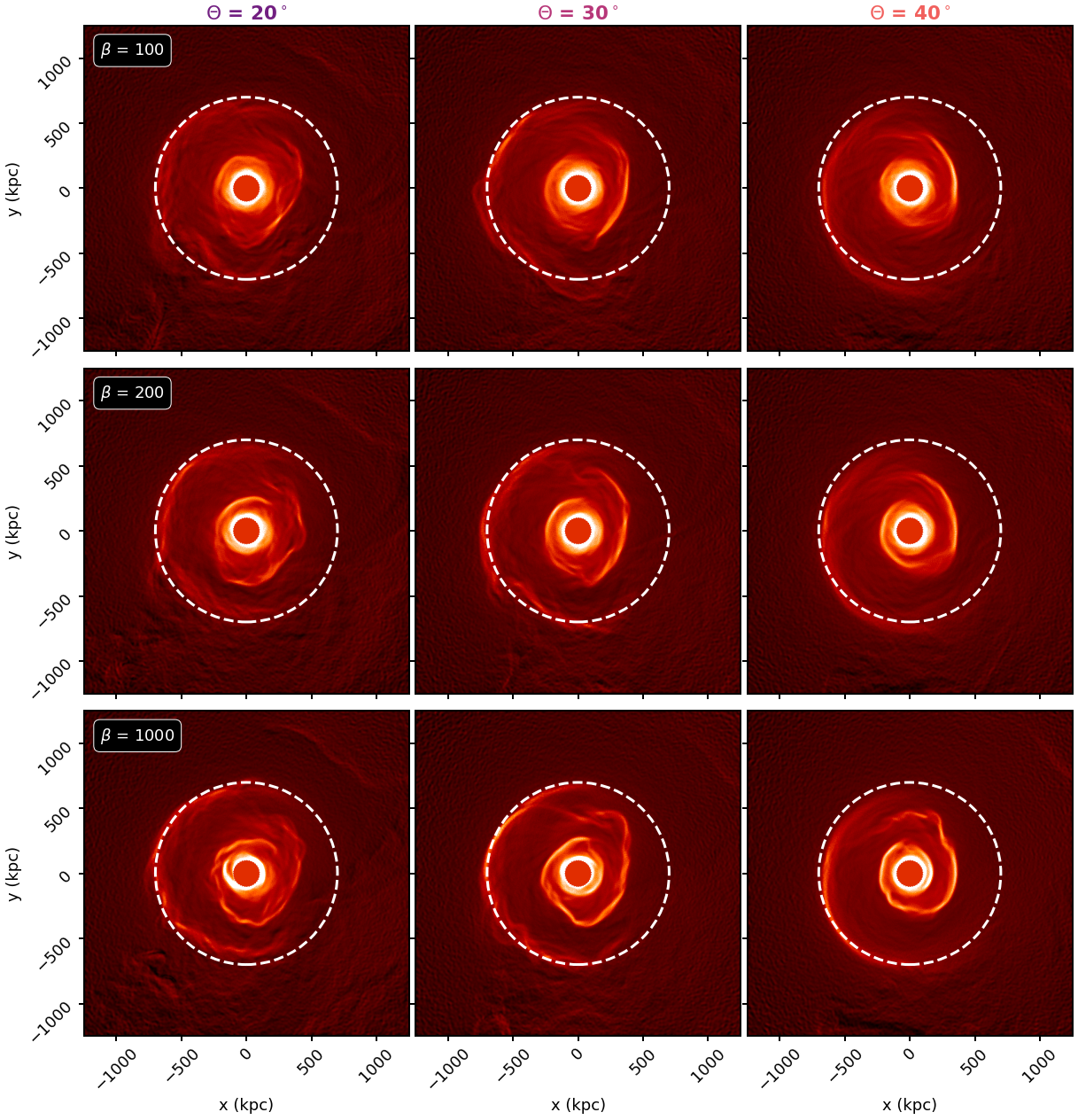}
        \caption{\textit{Different $\theta$ and $\beta$.} GGM images showing the edges of the SB for simulations of mergers with a gaseous subcluster with mass ratio $R$~=~1:5, initial velocity $v_{\rm sub}= 2000$ km~s$^{-1}$ and for different values of $\beta$ = 100, 200, 1000  and $\theta =20^\circ, 30^\circ, 40^\circ$. The dashed line in each panel marks a circle of radius 700 kpc. All images are shown at 7.5 Gyr after the first pericenter passage.}
        \label{fig:simu_compare_theta_beta_GGM}
        
    \end{figure*}
    \begin{figure*}[!t]% Fig.9
        \centering
        \includegraphics[width=0.9\textwidth]{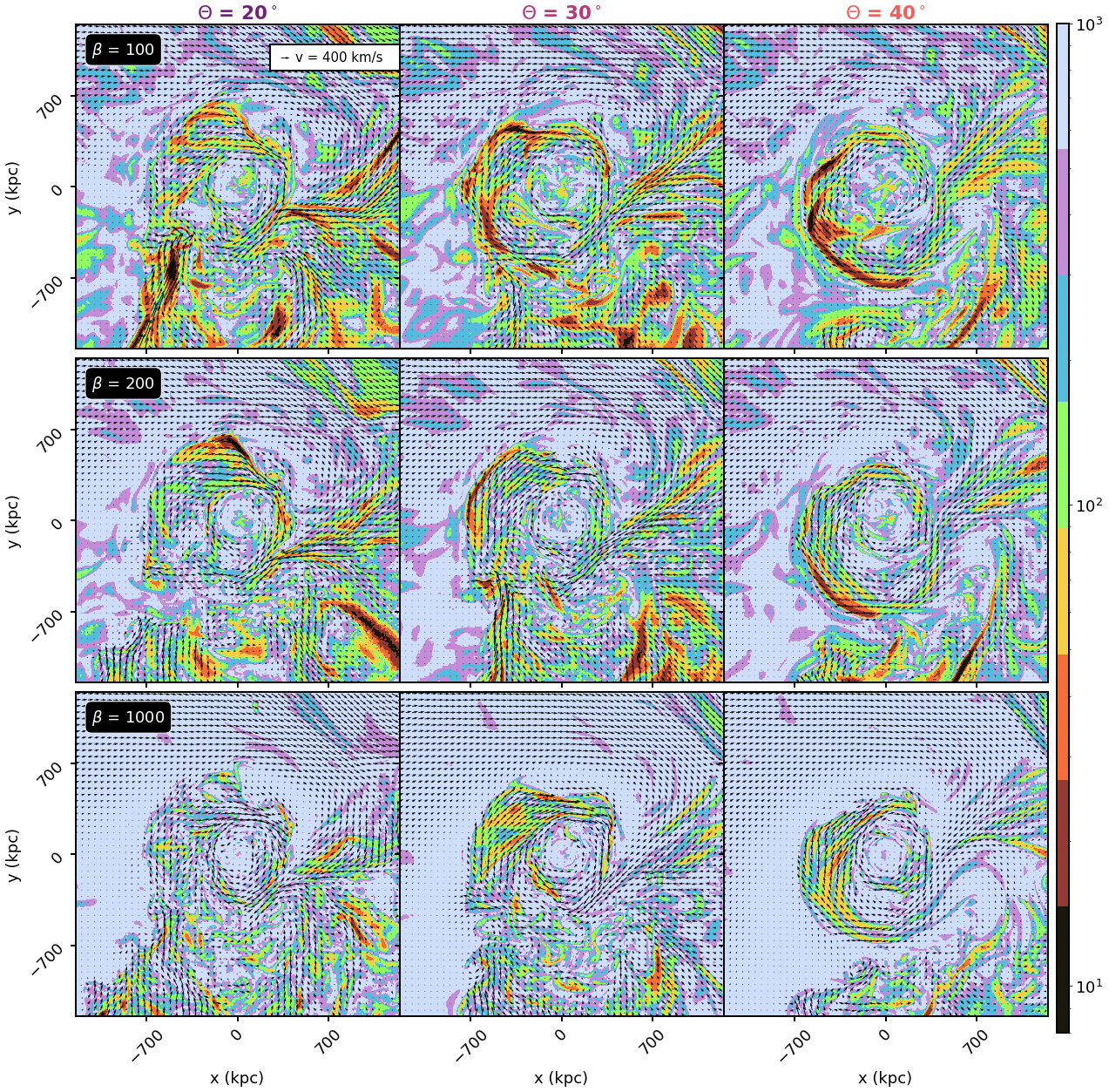}
        \caption{\textit{Different $\theta$ and $\beta$.} Slices of the local $\beta$ values for simulations with a gaseous subcluster with mass ratio $R$~=~1:5, initial velocity $v_{\rm sub}= 2000$ km~s$^{-1}$, and for different initial $\beta$ and $\theta$ values. The slices are taking 7.5 Gyr after pericenter passage, in the $x-y$ plane of the simulation (the merger plane) and centered on the potential minimum of the main cluster. Arrows indicate the velocity of the gas particles in both direction and magnitude. All images are shown at 7.5 Gyr after the first pericenter passage.}
        \label{fig:simu_compare_theta_beta_quiver}
        
    \end{figure*}   
    
    % #############################################
    % Effect of BETA               
    % #############################################
    \subsection{Effect of the initial magnetic field}\label{sec:effect_beta}
    %\subsection{Impact of the magnetic field}
        %\begin{figure*}
        %    \centering
        %    \includegraphics[width=1\textwidth]{CF_GGM_cut_gas_theta30_R02_vel2000.png}\\
        %    \includegraphics[width=1\textwidth]{CF_beta_cut_gas_theta30_R02_vel2000.png}
        %    \caption{Cut in the GGM of the est CF showing the effect of $\beta$ on the smoothing of the cold front. The simulations shown here are R=1:5, $v_{\rm sub}= 2000$ km~s$^{-1}$, $\theta =30^\circ$.
        %    }
        %    \label{fig:CF_cut}
        %\end{figure*}
        
    In this section, we analyze the critical role of the magnetic field (here parametrized by the plasma parameter $\beta$) on the evolution of the cold front properties.

    %We now detail the effect of varying the initial magnetic field strength (or, equivalently, plasma $\beta$ parameter) on the properties of the cold fronts. %, as initially investigated for a different cluster setup in \citet{zuhone_sloshing_2011}.

    %%% GGM
    In Fig.~\ref{fig:simu_compare_theta_beta_GGM}, we show the GGM image of the SB of different simulations at 7.5 Gyr after pericenter passage. Each column in the figure corresponds to a different initial angle, and each row corresponds to a different magnetic field strength, going from a strong initial magnetic field ($\beta = 100$) to a weak one ($\beta = 1000$). As noted in previous investigations \citep{zuhone_sloshing_2011}, the magnetic field has the effect of smoothing the CFs from the effects of KHIs. In the absence of strong fields, KHIs can develop and CFs have a more ``boxy'' shape overall. Needless to say, the ability of even strong magnetic fields to keep cold fronts smooth is limited if the subcluster can strongly perturb the gas initially and at multiple subsequent core passages (especially if it is gaseous), which is seen especially in the case with $\theta = 20^\circ$ (left column).\footnote{In their investigations of sloshing cold fronts in the presence of a magnetic field, \citet{zuhone_sloshing_2011} did not vary the mass ratio or the trajectory of the subcluster, but only the magnetic field strength and its spatial distribution.}  

    %%% beta
    The same snapshots plotted in Fig.~\ref{fig:simu_compare_theta_beta_GGM} are shown in Fig. \ref{fig:simu_compare_theta_beta_quiver}, where we plot the plasma parameter $\beta$  in a slice through the merger plane centered on the main cluster's center at 7.5 Gyr after pericenter passage. In general, the magnetic field is strongly amplified by the sloshing motions (with the plasma $\beta$ decreasing to $< 10$ in the most magnetized layers), as was already pointed out by \citet{zuhone_sloshing_2011}. This amplification mainly happens inside the cold fronts, where the shear is larger, but there is also significant amplification of magnetic field along gas streams and flows outside of the fronts. 

    \subsection{Cold Fronts Expansion Speed}\label{sec:expansion_speed}
        \begin{figure*}
            \centering
            \includegraphics[width=1\textwidth]{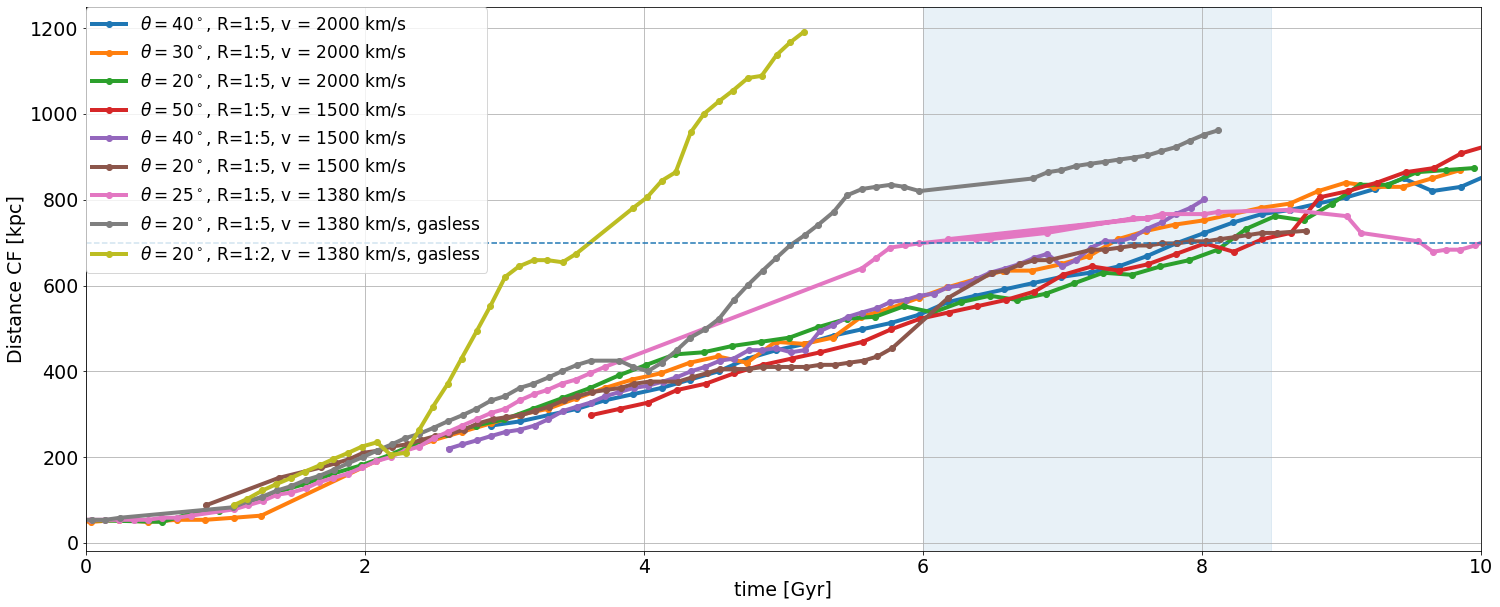}
            \caption{\textit{Cold Front Expansion Speed.} Plots of the first CF position as a function of time for different simulations, showing the distance traveled by the CF over time. The blue horizontal line indicates the observed position of the ancient CF in Perseus, and the blue area indicates the time span between 6 Gyr and 8.5 Gyr, the time range at which the initial CF in the different simulations reaches the radius of $\sim$700~kpc. 
            }
            \label{fig:CF_speed}
        \end{figure*}

    The ``ancient'' cold front at $r \sim $~700~kpc is produced in a number of our simulations. 
    In Fig~\ref{fig:CF_speed}, we show the position on the first cold front as function of time from the pericenter passage. 
    The time at which the first CF reaches $r \sim $~700~kpc (see horizontal dashed blue line) from the center generally depends on the initial parameters, largely (as we describe below) if the combination of parameters results in multiple pericenter passages.
    In most cases shown, the CF has an almost constant speed of $\sim$100~km~s$^{-1}$ \citep[the constant expansion speed of sloshing cold fronts was also noted in previous works, e.g.][]{roediger_gas_2011,Roediger2012}. For simulations with one core passage, the expansion speed of the CF is relatively independent of the initial parameters of the subcluster (provided it is massive enough to produce fronts in the first place), and the CF reaches the $\sim$700~kpc radius in $\sim$6-8.5~Gyr after pericenter passage (as indicated by the light blue area). 
    
    In the case of multiple passages, the subcluster generally returns before the CF reaches this radius, which can disrupt it and change its velocity. 
    The yellow and gray lines in Fig.~\ref{fig:CF_speed} show this scenario clearly. In these cases the subcluster is more massive and/or the impact parameter small, the gas of the Perseus-like cluster is perturbed multiple times, and the first CF reaches the radial distance of 700 kpc before 6 Gyr. In the gaseous cases, the CFs are more disrupted by the turbulence and the repeated passages of the subcluster, so in Fig~\ref{fig:CF_speed} we show the gasless cases, in which the CFs are more visible and easier to track (see Fig.~\ref{fig:simu_compare_MASS_multiplepassages_GGM} to compare the appearance of the CFs in the gaseous and gasless cases).

    There are also intermediate cases, in which the subcluster undergoes multiple passages, but the additional perturbations on the CFs are not strong enough to radially accelerate them in a strongly visible way. These cases are where the impact parameter is larger (pink curve) or the velocity is larger (brown curve). In these cases, as in the case of a single passage, the most probable age of the ancient CF is around 7-8 Gyr.
    
    %pushing it and changing its shape. This can sometimes lead to the disruption of the CF. The expansion of the CF is modified by the passage of the subcluster or the shock waves and fronts that these subsequent passages can produce.

    %When the subcluster comes back the expansion of the cold front is affected by the subcluster movement and the subsequent shock fronts. 
    
    %Generally, the CF has an almost constant speed of $\sim$100~km~s$^{-1}$ unless it is overrun by a shock front. %, in which case the CF stops and then starts again its expansion, usually faster than before.  On the contrary, 
    %Explain the two outliers. One is R = 1:2, which probably explains why it is different. What is different about the R = 1:5, gasless one? presumably it’s the low impact parameter from the theta = 20.
    However, in the case of multiple subcluster passages, the more frequent perturbations of the gas affect the radial evolution of the CFs.
    If the CF is not destroyed by the multiple passages and the shock, it reaches the 700~kpc radius in a shorter timescale (see the pink and brown lines in Fig.~\ref{fig:CF_speed}) because of the multiple kicks that the gas receives. The larger the mass of the subcluster, the larger the kicks, hence, the larger the expansion speed of the CFs.

    \section{Discussion}\label{sec:discussion}
    
    % -------------------------
    %        Hot Split                      
    % -------------------------
    \subsection{Hot split in CF}\label{sec:split}

        \begin{figure*}
            \centering
            \includegraphics[width=1\textwidth]{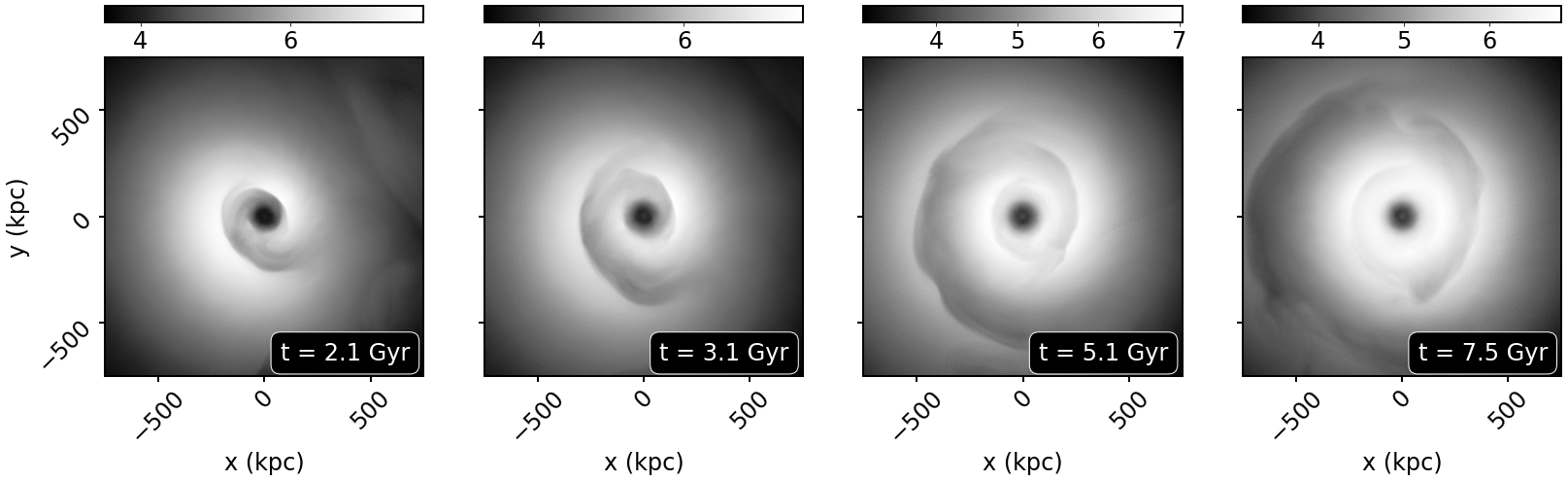}
            \caption{\textit{Evolution of the split.} Projected temperature maps for simulations of mergers with a gaseous subcluster with mass ratio $R$~=~1:5, initial velocity $v_{\rm sub}= 2000$ km~s$^{-1}$ and for different values of $\beta$ = 200  and $\theta =30^\circ$. The time indicated in the bottom right corner of each image is from the first passage. The temperatures are indicated in keV.
            }
            \label{fig:hook_evolution}
        \end{figure*}
        
    \begin{figure*} %Fig.12
            \centering            
            \includegraphics[width=1\textwidth]{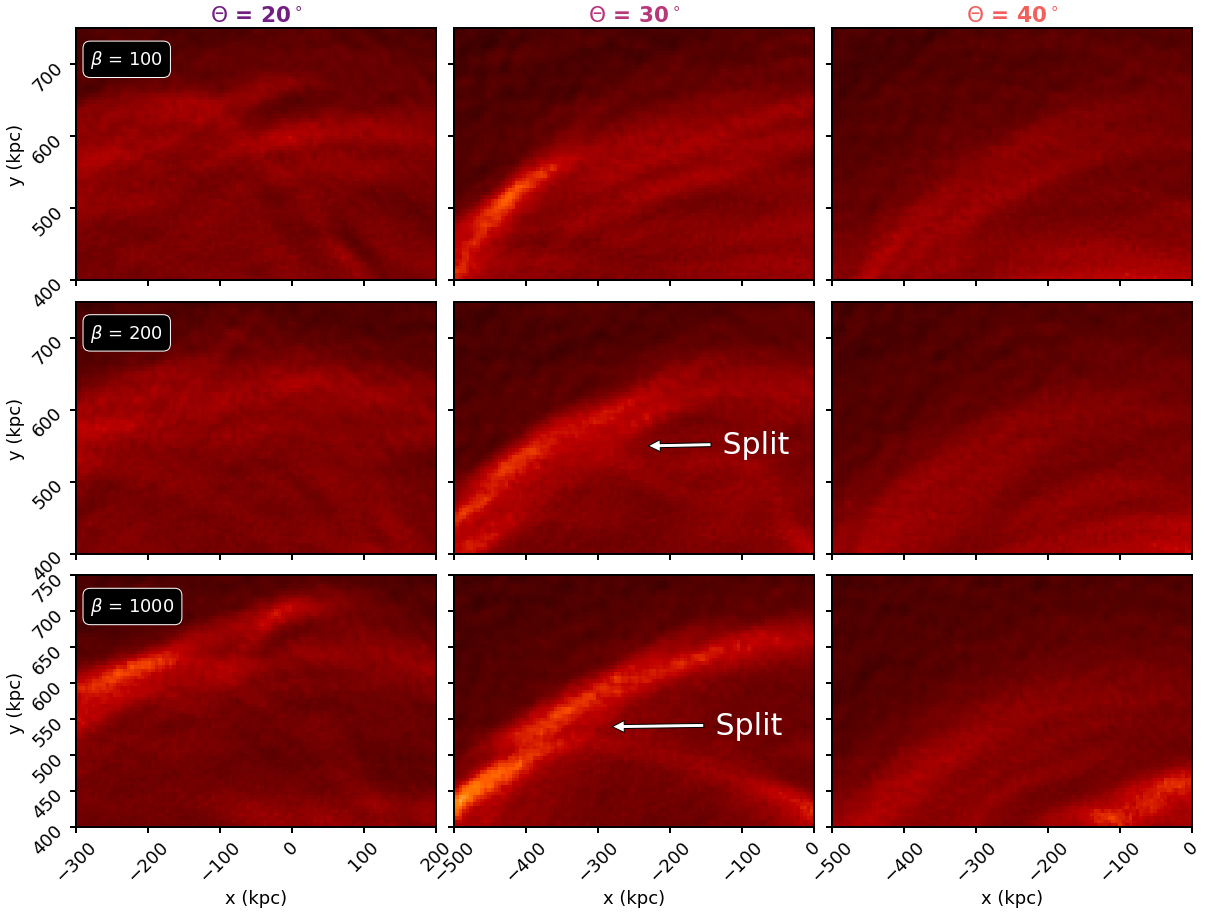}
            \caption{\textit{Different $\theta$ and $\beta$.} GGM maps for simulations of mergers with a gaseous subcluster with mass ratio $R$~=~1:5, initial velocity $v_{\rm sub}= 2000$ km~s$^{-1}$ and for different values of $\beta$ = 100, 200, 1000  and $\theta =20^\circ, 30^\circ, 40^\circ$.  All images are shown at 7.5 Gyr after the first pericenter passage. The split is marked by the white arrow.  
            }
            \label{fig:GGM_zoom_split}
        \end{figure*}
    
    %%% general: split, bifurcation of the CF + hot tongue
    In most of the GGM images shown above, there is a split in the northern part of the outer CFs, similar to the one seen by \citet{walker_split_2018}. Such splits are produced in simulations of sloshing shortly after a CF forms; some of the gas in the front is of lower entropy and falls back towards the center, while the rest of the gas continues onward, which produces a split between the two.

    %%% our simu
    Previously, in Fig. \ref{fig:simu_compare_MASS_singlepassage_slices}, we showed temperature slices (second column) for three simulations. A split in the CFs is apparent in both the $R$~=~1:5 gaseous and gasless cases (middle and bottom panels of the second column).  Inside the split, a hotter tongue of gas appears in between the fronts. The same figure shows that this hot tongue has a tangential flow into the hook going in the opposite direction from the flow underneath the front surface (fourth column). 

    In Fig.~\ref{fig:hook_evolution} we plot the evolution of the split in the projected temperature for one of the simulations. 
    The split appears at an earlier epoch and then it grows as the CFs expand. The projections of temperature have the effect of smoothing out the split which appears less drastic than in the temperature slices. Despite this, the split is still visible.
    
    %In Fig. \ref{fig:proj_temp_compare}, we plot the projected emission-weighted temperature maps for some of the simulations, computed using \cite{mazzotta_comparing_2004} weighting. 
    In Fig.~\ref{fig:GGM_zoom_split}, we zoom in on the nominal location of the split of the external CF in several simulations with varying magnetic field and initial angle of the subcluster $\theta$.  The azimuthal position of the split is highly dependent on the initial angle of the subcluster (see also Fig. \ref{fig:simu_compare_theta_beta_GGM}, for small impact angles the split is located north, while for larger impact angles, the split is located more north-east. For this figure, we orient the split in the same direction in all simulations. 

    The split is present to varying degrees in the simulations. With a small impact parameter (left column), the CF split is disturbed by the turbulence driven by the rapid return of the subcluster to the core region, and would likely not be identified as a split, with the possible exception of the $\theta = 20^\circ$, $\beta = 1000$ case (bottom-left panel). The presence of a split is most clear when $\theta = 30^\circ$ (center column). When $\theta = 40^\circ$, a faint hint of a split is present, but it is more difficult to identify. With such a large impact parameter, less momentum is transferred to the core gas, resulting in a less developed contrast.

    The appearance of the split is qualitatively similar for $\beta = 1000$ (bottom row) and $\beta = 200$ (middle row), but for a strong initial magnetic field ($\beta = 100$), the enhanced magnetic pressure smooths out the split (top row) and makes the contrast less obvious. 

    These results are broadly consistent with that of \citet{walker_split_2018}, who first presented simulations of a CF split along with the first observations of a split in Perseus, and showed a similar dependence in appearance on the magnetic field strength.
    
    % -------------------------
    %   Additional Features + 1.2-1.7 Mpc fronts        
    % -------------------------
    \subsection{Edges at larger radii}
    \begin{figure*}
        \centering
        \includegraphics[width=1\textwidth]{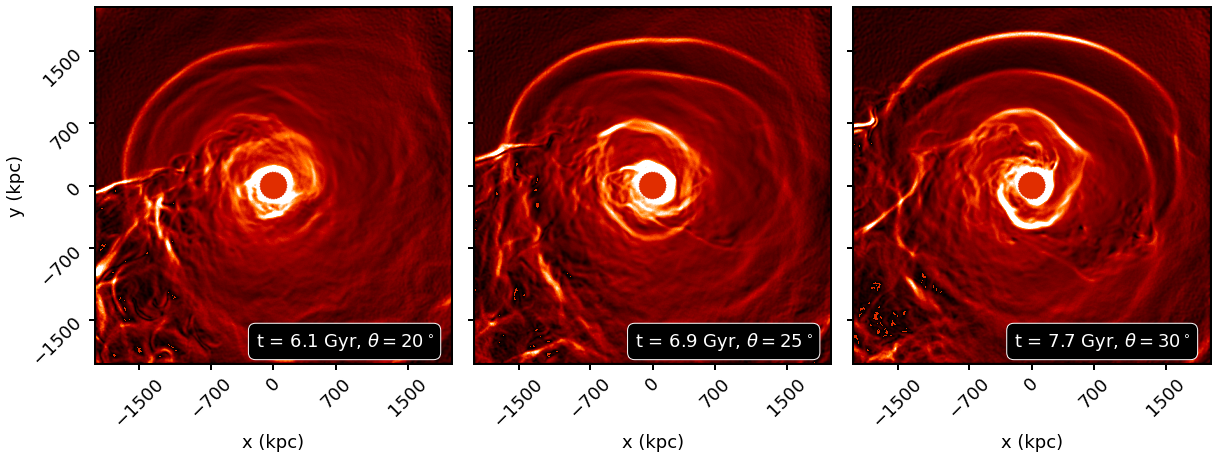}
        \caption{GGM images for different projection angles for the simulation with $\beta=200$, $R$~=~1:5, showing the external edges. All images are shown at 7.5 Gyr after the first pericenter passage. 
        }
        \label{fig:GGM_externaledges}
    \end{figure*}
    
    Other SB edges at large radii of $\sim$1.2 and 1.7~Mpc were observed in the Perseus Cluster by \citet{walker_is_2022} with \textit{XMM-Newton} (the existence of the edge at 1.7~Mpc is less certain due to possible stray light effects). Using \textit{Suzaku} data, they showed temperature jumps at these edges consistent with CFs, though the errors are large due to the faintness of the features. Confirmation of the precise nature of these discontinuities will require follow-up observations.

    In many of our simulations, edge features can indeed appear at such large radii, especially if there are multiple core passages, which drive multiple shocks into the ICM which expand to large radii and form concentric SB edges. Because the shocks travel outward at a speed an order of magnitude higher than the CFs, they can travel out to radii of 1.7 Mpc in less than $\sim$2~Gyr from their creation in the core region. By contrast, in most of our simulations, sloshing CFs take more than the age of the universe to reach these radii. There are some rare multi-passage cases, with a high mass ratio / large subcluster, in which CFs can expand to 1.2~Mpc, but never to 1.7~Mpc.

    Double SB edges at large radii appear in nearly all of the simulations with multiple core passages, and in some cases separated by $\sim$ 500 kpc as in the observations. 
    Not only does the location of the double SB edges depend on the initial angles of the incoming subcluster, but also their separation (e.g., the separation in $\sim$500 kpc for the case $\theta = 30^{\circ}$, but smaller for the smaller initial angles.)
    However, in our simulations, we are unable to obtain a qualitative match with the Perseus observations in terms of the azimuthal angle between these fronts and the CF at 700~kpc, which depends in general on the trajectory of the subcluster. It is also important to note that our idealized simulations do not include the effects of cosmological accretion, and therefore there is no accretion shock in the outskirts that would exist in real systems. As shown by \citet{zhang_2020_,zhang_2020}, the two edges near the virial radius could be produced by the collision of a merger shock with an accretion shock. While these fronts would not be reproducible with our current setup, the \cite{zhang_2020} study of the growth of a Perseus-like cluster shows that the gas entropy profiles within $r \sim$1~Mpc are largely stable for the last $\sim$7~Gyr, hence the region studied in this work ($r \lesssim 700$~kpc) is not likely to be strongly affected by the accretion shock and its interaction with the ``runaway'' shocks created by the subcluster. % probably multiple pericenter passages would help in keeping the inner entropy profile for even longer

    As already noted by \cite{birnboim_2010}, when a shock wave passes through a cold front, it can cause the front to expand and become less sharp. In some cases, the shock can completely destroy the cold front and cause the gas to mix together.
    The strength and orientation of the shock wave are also important factors. If the shock wave is strong and moves perpendicular to the cold front, it can cause significant expansion and mixing, it can disrupt the cold front itself or create enough turbulence to split the CF into multiple segments. The shock front is stronger for a higher mass ratio. When the shock is weaker or moves parallel to the front, the effects are less pronounced. As an aside, the study by \cite{birnboim_2010} showed that two colliding shocks can give rise to CFs, generally quasi-spherical, which can happen at large radii. In our simulations, shocks from later pericenter passages never overtake earlier ones, and such CFs are never produced. 
    
    There are few epochs in the multi-passage simulations where we get double edges. In Fig.~\ref{fig:GGM_externaledges}, we show three of them for different projection angles for the simulation with $\beta=200$, $R$~=~1:5. These double external edges are all expanding shock fronts.

    % -------------------------
    % Projections effects                  
    % -------------------------  
    \subsection{Projection Effects}

    \begin{figure*}
        \centering
        \includegraphics[width=1\textwidth]{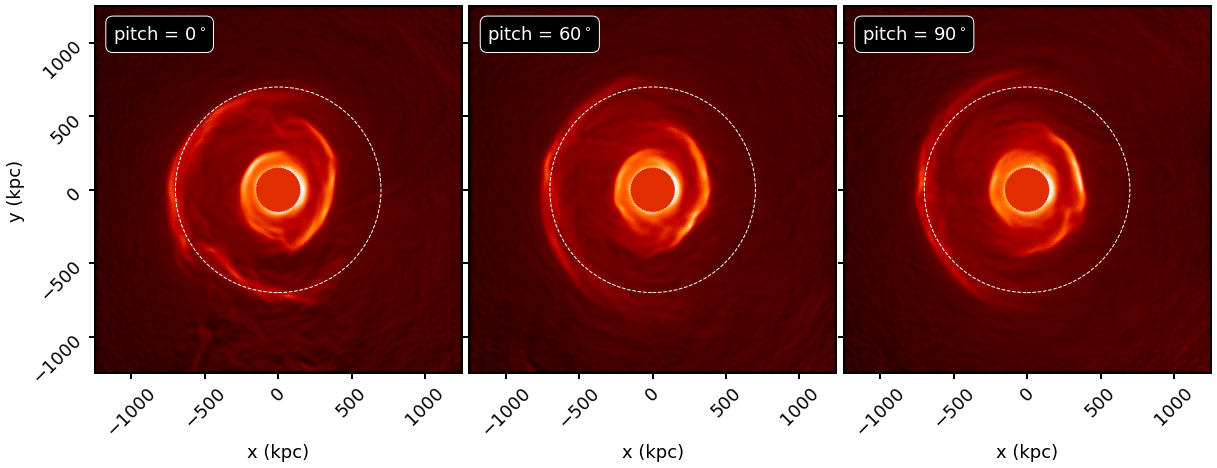}\\
        \includegraphics[width=1\textwidth]{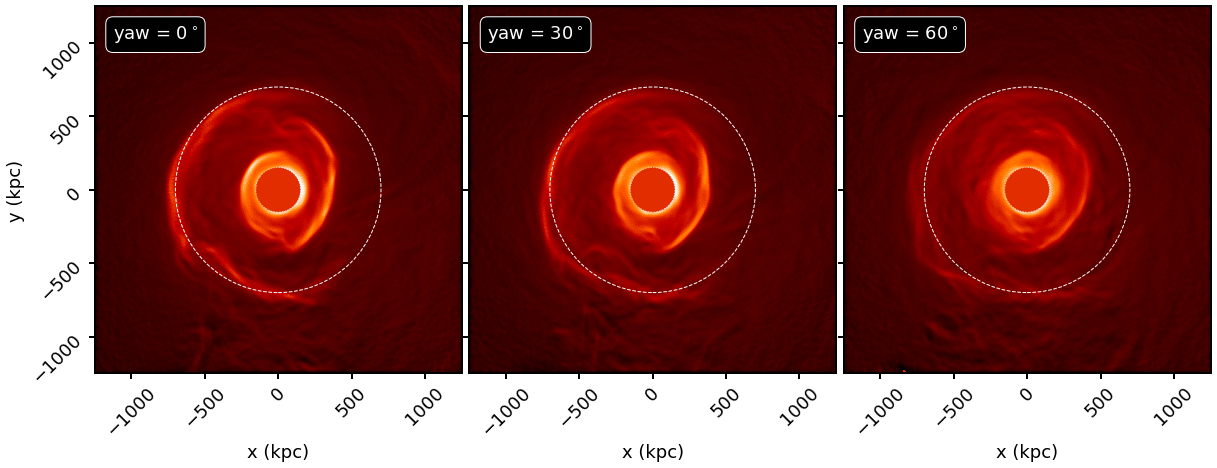}
        \caption{GGM images for different projection angles for the simulation with $\beta=200$, $R$~=~1:5, $\theta = 30^{\circ}$ and $v = 2000$ km~s$^{-1}$. The dashed line in each panel marks a circle of radius 700 kpc. All images are shown at 7.5 Gyr after the first pericenter passage. 
        }
        \label{fig:GGM_pitch}
    \end{figure*}
    
    In all the previous discussions, the analysis has been done assuming that the merger happened in the plane of the sky. Here, we briefly discuss the effect of observing our simulations at an angle between the plane of the merger and the plane of the sky.
    In Fig.~\ref{fig:GGM_pitch}, we show the off-axis projections with different pitch and yaw angles for one of the simulations. Here, we chose to show the simulation with  $\beta=200$, $R$~=~1:5, $\theta = 30^{\circ}$ and $v = 2000$ km~s$^{-1}$.
    For a merger happening in the $xy$ plane, the pitch angle is the angle from the $z$ axis in the $yz$ plane, and the yaw angle the one from the $z$ axis in the $xz$ plane. When both these angles are zero, the line of sight corresponds to the $z$-axis of the simulation.
    In the normal plane-of-sky projection the CFs in this simulation have a ``boxy'' appearance, showing KHI when the plane of the merger is in the plane of the sky. However, these features are less evident when increasing the angle between the line of sight and the merger plane. 
    Moving in the $yz$-plane or in the $xz$-plane has very different effects.
    When increasing the pitch angle, the angle of the split in the outermost CF increases, and for a pitch angle of 90$^\circ$ (looking into the merger plane) the inner and outer CF appear completely disconnected.
    Increasing the yaw angle and moving in the $xz$-plane, the CF gradients appear less sharp, as we are projecting along a direction where different parts of the CFs have different radii of curvature, smearing the edge out when viewed in projection. The same differences in the radii of curvature also result in slight differences in the observed projected radii of the CFs. 
    %The CF is not actually a surface (une calotte) so when we move in a direction "parallel" to this surface - in the previous
    
    %This is true if we move at an angle
    
    %While this is true when moving in the $yz$-plane - the plane of the merger is defined as $xy$ - when moving in the $xz$-plane the projections make the CFs "messy". %that lose their structure and contrast. % makes sense, due to the geometry of the merger

    %In Fig.~\ref{fig:GGM_pitch} we can see the split with other projection angles. For a wide range of projection angles, this split is still visible. When then the angle increases too much the effect of the projection prevails and the split disappears.
    %We also point out that the precise position of the CF has a slight dependence on the projection angle, since the surfaces in general are not sections of spheres but instead have varying radii of curvature in different directions. 

    % -------------------------
    % Donut               
    % -------------------------  
    \subsection{Evolution of CFs in the core region}\label{sec:innerCF}
    \begin{figure*}
        \centering
        \includegraphics[width=1\textwidth]{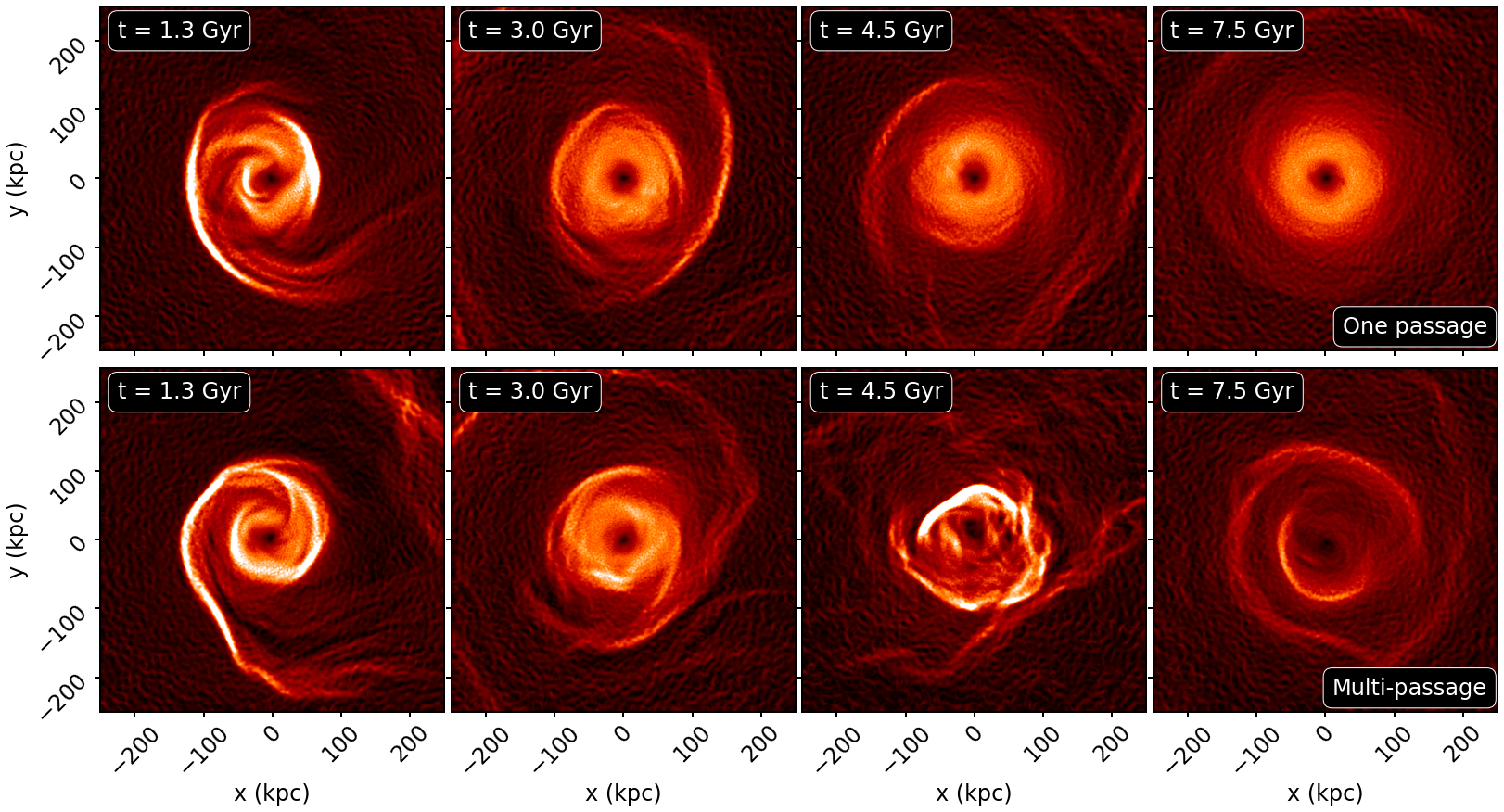}
        \caption{GGM images of the core region for the simulation with $\beta=200$, $R$~=~1:5, $\theta = 20^{\circ}$ and $v = 2000$ km~s$^{-1}$. The time indicated in the top left corner of each image is from the first passage. 
        }
        \label{fig:innerCF_singlepassage}
        
    \end{figure*}

    Throughout this work, we have so far focused on the outermost CF, and have masked the inner region in the GGM images (e.g., Fig.~\ref{fig:simu_compare_MASS_multiplepassages_GGM}, \ref{fig:simu_compare_MASS_singlepassage_GGM}, \ref{fig:simu_compare_theta_beta_GGM}). In Fig.~\ref{fig:innerCF_singlepassage}, we show the evolution of the inner CFs. In the top panels, we show the GGM images in the core region of a single-passage merger. The first CF forms after the first core passage, and by $\sim$4~Gyr after this time it leaves the core region. With the passage of time, the core region progressively loses the spiral structure with time, the sloshing motions are less pronounced, and the inner CFs increasingly lose their distinctness. 
    This process is partially due to the mixing of cold gas with warmer gas from larger radii, which has the effect of flattening out the temperature and the entropy profiles in the central region \citep[reducing the entropy contrast needed to produce a CF in the first place, see][]{zuhone_stirring_2010}, as well as the damping of the original gas motions in the core. 

    By contrast, if a subsequent passage does occur (bottom panels), a new round of sloshing motions begins in the core, producing new CFs. Due to the flattening of the entropy in the core region, the CF gradients are not as large. In real clusters, this entropy flattening would be mitigated somewhat by radiative cooling of the gas; we leave the inclusion of this effect in our simulations for future work. 
    
    Thus, to produce CFs at smaller radii in addition to the older fronts which appear at larger radii, a second perturbation by a subcluster appears to be necessary. Nevertheless, this does not imply a second passage by the same subcluster, as shown here. The second perturbation could be produced by a second subcluster; such scenarios were modeled by \citet{Vaezzadeh2022}. If an observed set of CFs at specific radii and azimuthal angles cannot be reproduced by a two-body merger, such a three-body interaction may need to be considered.

    \subsection{Comparison with Observations}
    \begin{figure*}
        \centering
        \includegraphics[width=0.8\textwidth]{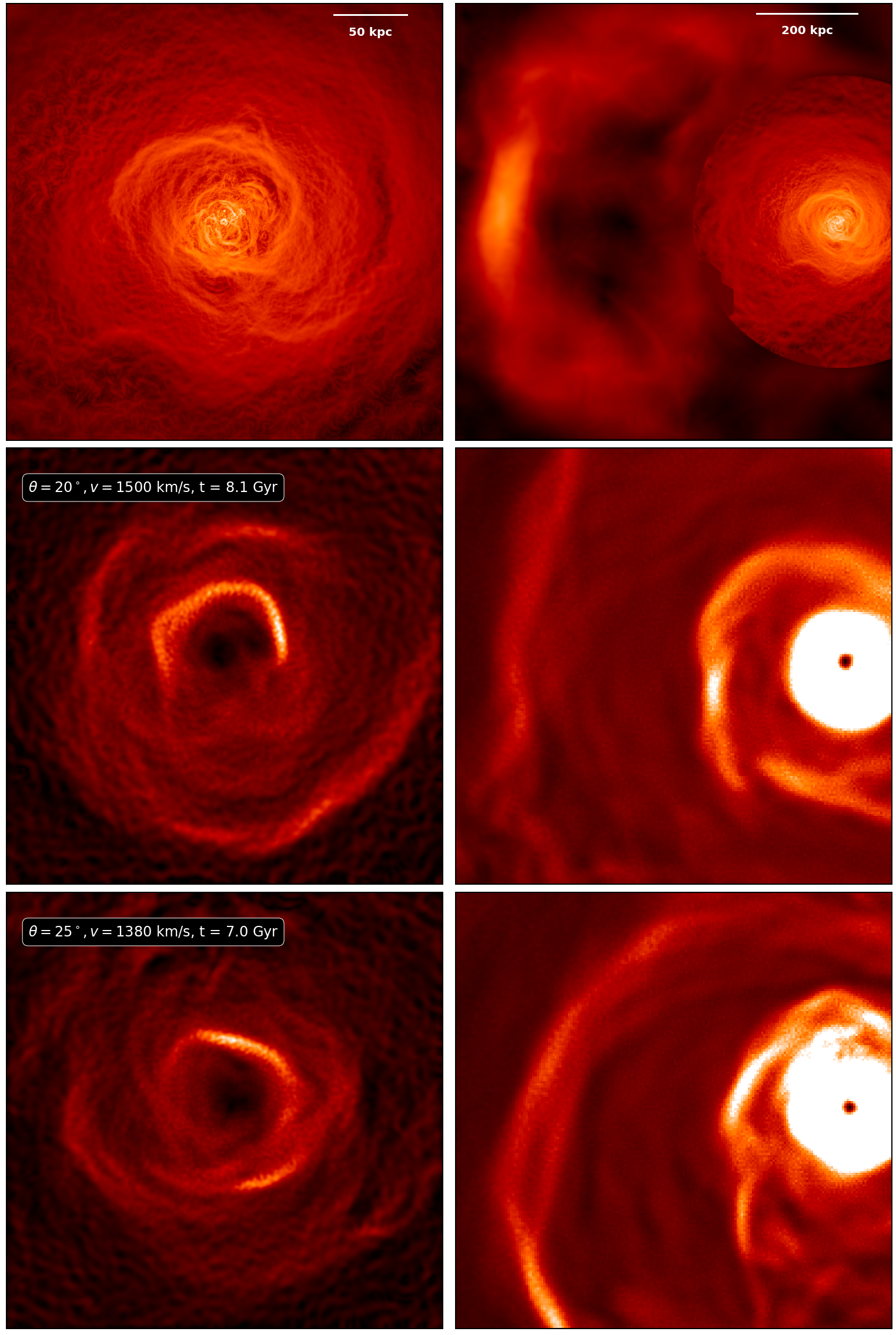}
        \caption{Comparison of the observations of Perseus \citet{walker_split_2018} (top row) with the results of two of our simulations (central and bottom rows). The initial parameters of the two simulations are indicated on the top left corner of the row corresponding to the simulation. All images show GGM maps. The images have been rotated by 90$^\circ$ to better match the observations.
        }
        \label{fig:GGM_bestagreement}
    \end{figure*}

    Finally, we attempt to identify the simulations that provide the best quantitative and qualitative agreement with the observations of the Perseus Cluster. Our criteria for a ``good match'' are the presence of a cold front that reaches a radius of 700~kpc, as well as cold fronts in the core region within $\sim$50-100~kpc. Since there is no obvious observed subcluster candidate in Perseus, the perturbing subcluster should either have made only one core passage or should have already completely merged with Perseus.
 
    We find two simulations that provide a close match under these conditions: ($R = 1/5$; $\theta = 20^\circ$; $v_{\rm sub} = 1500$), and ($R = 1/5$; $\theta = 25^\circ$; $v_{\rm sub} = 1380$). In both cases, the subcluster has merged into Perseus before the end of the simulation. In the former case, the appropriate epoch is 8.1~Gyr after the first core passage, and in the second it is 7.0~Gyr. In Fig.~\ref{fig:GGM_bestagreement} we compare the observed GGM images of Perseus from \citet{swal16} and \citet{walker_split_2018} with the same maps from these two best matches from our simulations, where the left panels show the inner core region and the right panel show the CF at a radius of 700~kpc. Though there is good agreement of the radial positions of the CFs in the simulations and observations, there is less agreement of the azimuthal phase of the fronts, especially in the core region. This particular feature may be difficult to reproduce, especially since in these simulations we have ignored physics that will be important for the ongoing evolution of the core region, such as AGN feedback and radiative cooling, which we reserve for future work. 

    % notes
    %It is interesting to notice that these two simulations show an outer front at 1.2 Mpc that could match the observations.
    % -------------------------
    % Degeneracy             
    % -------------------------  
    %\subsection{**Degeneracy}
    %It is interesting to notice that some very different initial conditions can give raise to similar CF. For example, the increasing incident angle of the incoming subcluster has a similar effect to the increasing magnetic field. As we can notice in Fig.\ref{fig:simu_compare_theta_beta}, both have the effect of smoothing the CF.

    %In order to have CFs in the inner region, a second perturbation of the core is necessary. With no second perturbation the inner CFs do not appear and the inner part appear as a donut in the GGM images.
    %The fact that a second perturbation is necessary to match the observations, does not imply that the second perturbations needs to be a second passage of the same subcluster, but it could be a different subcluster. 
   
%%%%%%%%%%%%%%%%%%%%%%%%%%%%%%%%%%%%%%%%%%%%%%%%%%%%%%%%%%%%%%%
%%%     CONCLUSION                      
%%%%%%%%%%%%%%%%%%%%%%%%%%%%%%%%%%%%%%%%%%%%%%%%%%%%%%%%%%%%%%%
\section{Conclusion}\label{sec:conclusions}

        Sloshing cold fronts are the signatures of merger activity in the cores of galaxy clusters. One of the best-studied CF systems is the bright and nearby Perseus Cluster, with CFs in the core region (within $\sim$200~kpc), an ancient CF located at $\sim$700~kpc, and two possible fronts at larger radii of $\sim$1.2 and 1.7 Mpc.

        %provide valuable information on the physical processes that govern galaxy clusters, such as the acceleration of gas bulk velocity, the development of plasma instabilities, and the characteristics of magnetic fields.

        In this work, we used magnetohydrodynamical simulations of a large, cool-core cluster similar to Perseus merging with subclusters to attempt to reproduce the general characteristics of the Perseus CFs. We perform a parameter space investigation over a range of mass ratios, impact parameters, initial velocities, and subcluster gas content. Our major findings are as follows:

        \begin{itemize}
        
        %%% MASS and GAS
        \item The mass and the content of gas can have a very big effect on the CFs existence, shape, and position, especially in cases where the subcluster makes multiple core passages (see Fig.~\ref{fig:simu_compare_MASS_multiplepassages_GGM}). The gradient of the SB increases with the mass of the subcluster; the larger the mass, the bigger the jumps in the SB and the more visible the CFs appear. Our simulations show that low mass ratio encounters do not create visible CFs that can reach large radii. Hence, it is possible to already place strong constraints on some parameters of the merger, simply by the mere existence of a CF located at a distance of half the virial radius from the center.
        %%% gas
        \item When the subcluster is gaseous (instead of a purely gravitational perturbation often used in previous simulations), it more strongly accelerates the gas in the main cluster core, producing stronger jumps in density and temperature, driving higher tangential velocities in the sloshing gas, and creating more turbulent motions (see Fig.~\ref{fig:simu_compare_MASS_singlepassage_slices}).
        In our simulations, the presence of gas in the subcluster produces other observable features, such as a trail of infalling gas, and much more disturbed CF, consistent with  \cite{ascasibar_origin_2006}. This gas stream can trigger KHI on the CFs and can create a ``bay'' in the southern part of the outer CF.
        
        %%% one passage
        \item The initial impact parameter and initial velocity of the subcluster also impact the appearance of the CFs. These parameters determine the number and frequency of core passages, as well as how long the subcluster spends near the main cluster center and how long it takes for the merger to be completed. For smaller initial speeds, the pericenter distance is smaller, the gravitational perturbation is greater, and since the subcluster is more gravitationally bound it perturbs the gas more frequently, affecting the evolution of the CFs.
        
        \item Simulations with one core passage produce relatively smooth and undisturbed CFs due to the absence of perturbations from subsequent passages (see Fig.~\ref{fig:simu_compare_MASS_singlepassage_GGM}). In contrast, if the subcluster returns for multiple core passages, the CFs appear more disrupted due to the effects of turbulence and KHI, especially in the case of a gaseous subcluster, also reducing the SB gradients.
        
        \item However, if there is only one core passage, our simulations show that a single passage is unable to reproduce the inner CFs observed in Perseus along with the larger CF (see Fig.~\ref{fig:innerCF_singlepassage}). The production of these inner CFs at a later time requires multiple passages (if indeed only a single two-body interaction is considered). This indicates that there is a narrow range of parameters around $R \sim 1:5$, $v \sim 1500$~km~s$^{-1}$, and $\theta \sim 25-30^\circ$, which are capable of qualitatively producing the main features we see in Perseus. Given these parameters, the age of the CF at 700~kpc in Perseus is between 7~Gyr and 8.5~Gyr from the first encounter with the subcluster. 

        \item The expansion speed of the initial CF is almost constant $\sim 100$ km~s$^{-1}$ unless the CF gets kicked by a shock front that has the overall effect of accelerating the expansion of the CF at larger radii. This is particularly visible in the case of a large number of passages, e.g., the subcluster is massive and/or the impact parameter small (see yellow and gray lines in Fig.~\ref{fig:CF_speed}). In the case of a single passage of the subcluster, the radial velocity of the CF is relatively independent of the initial parameters.
        
        %%% age of the cold front
        %\item From this work, the age of the CF at 700 kpc in Perseus is between 7 Gyr and 8.5 Gyr from the closest encounter/ passage of the subcluster. Depending on the initial conditions of the merger, the developed CFs show different characteristics and evolution. When a more massive subcluster kicks the gas of the main cluster, the first-formed CF reaches half of the virial radius earlier.

        %%% features at larger radii
        \item None of our simulations show cold fronts at radii larger than the critical radius $R_{\rm 200c}$, but they do produce shocks that can reach larger radii in the same time frame due to their faster speeds. These shock fronts are created in the case of multiple passages of the subcluster near the core region (e.g. Fig~\ref{fig:GGM_externaledges}).

        %%% split / tongue
        \item Some sloshing CFs simulated have a split in the northern part and a tangential velocity that goes from the larger part of the spiral inward, to the exception of a tongue of gas that is created sometimes by the bow shock in front of the incoming subcluster. This tongue is moving in the opposite direction, creating a split in the CF. This tongue is visible in both the GGM and the temperature maps, and it is characterized by a hotter temperature than the CF (see Figs.~\ref{fig:simu_compare_MASS_singlepassage_slices}, \ref{fig:hook_evolution},\ref{fig:GGM_zoom_split}). % if it has a righthand rotation, this give us constrains on the mouvement of the subcluster

        \item %From these simulations, the orientation of the spiral shape of the CFs gives us information on the direction of the incoming subcluster, i.e. a counterclockwise\footnote{counterclockwise if the direction of the spiral is defined from the inner part to the outer part} spiral shape corresponds to a counterclockwise movement of the subcluster. 
        Together, the direction of the tangential flow in the CF and the position of the tip of the split in the outer CF can give us information on where the subcluster came from. In the cases we identified as most closely matching the Perseus observations, the subcluster has already completely merged in with the main cluster by the time the CFs reach the radii they are observed at (see Fig.~\ref{fig:GGM_bestagreement}). 
        %- la spirale si avvita nel senso imposto dal passaggio del subcluster, quindi dall'avvitamente gia abbiamo una constrain sulla direzione del subcluster

        %%% cool-core
        %\item Perseus is a relaxed cluster, and our simulations show that the merger does not destroy the cool-core.
        % Perseus is a "relaxed" cluster, most of the initial conditions analyzed evolved in an ongoing merger in which the subcluster is still orbiting around the center of mass of the main cluster, losing progressively its mass.
        
        %\item As noticed in previous studies there is a degeneracy between certain parameters \cite{roediger_gas_2011}.
        
        \end{itemize}

        In future work, we will need to consider other physical effects in order to make our simulations more realistic. First, the large length of time between the first core passage and the observed position of the outermost CF is long enough that the outskirts of Perseus may significantly evolve due to cosmological accretion. The importance of this effect for our results will need to be considered. 
    
        %Future: apply Arithmetics to this parametric study
        
        %AGNs are responsible for the CF at small scales. How do AGNs influence the CF at a large scale? 
        In a subsequent paper, we will consider the effects of radiative cooling, accretion onto the central SMBH, and AGN feedback, and the interplay of these effects with the sloshing CFs in the core region.

\begin{acknowledgments}
EB acknowledges funding from Chandra grants TM1-22007X, GO1-22123C, and GO2-23122A, and Smithsonian Scholarly Studies grant. JAZ is funded by the Chandra X-ray Center, which is operated by the Smithsonian Astrophysical Observatory for and on behalf of NASA under contract NAS8-03060. The simulations were run on the Pleiades supercomputer at NASA/Ames Research Center. We would like to thank Bill Forman for useful comments, and acknowledge NASA Grant 80NSSC19K0116 and Chandra Grant GO9-20109X. RW acknowledges the funding of a Leibniz Junior Research Group (project number J131/2022). MR acknowledges the grant NSF AST-2009227. 
\end{acknowledgments}


\software{
    AREPO \citep{springel_e_2010},
    AstroPy \citep{2013A&A...558A..33A,2018AJ....156..123A},  
    Colossus \citep{diemer_colossus_2018},
    yt \citep{turk_yt:_2011} 
}

\bibliography{biblio}
%\begin{thebibliography}{}
%\bibitem[Ar{\'e}valo et al.(2016)]{are16} Ar{\'e}valo, P., Churazov, E., Zhuravleva, I., Forman, W.~R., \& Jones, C.\ 2016, \apj, 818, 14

%\end{thebibliography}

\end{document}